\documentclass[twocolumn,paper]{aa}

\usepackage{amsmath} 
\usepackage{amssymb}
\usepackage{graphicx}
\usepackage{subfigure}
\usepackage{natbib_copy}

\begin{document}
\title{Gravitational lensing in the Supernova Legacy Survey (SNLS).
\thanks{
Based on observations obtained with MegaPrime/MegaCam, a joint project
of CFHT and CEA/DAPNIA, at the Canada-France-Hawaii Telescope (CFHT)
which is operated by the National Research Council (NRC) of Canada,
the Institut National des Sciences de l'Univers of the Centre National
de la Recherche Scientifique (CNRS) of France, and the University of
Hawaii. This work is based in part on data products produced at the
Canadian Astronomy Data Centre as part of the Canada-France-Hawaii
Telescope Legacy Survey, a collaborative project of NRC and CNRS.
Based on observations 
obtained at the European Southern Observatory using the Very Large Telescope
on the Cerro Paranal (ESO Large Program 171.A-0486 \& 176.A-0589).
Based on observations (programs 
GS-2003B-Q-8,
GN-2003B-Q-9,
GS-2004A-Q-11,
GN-2004A-Q-19,
GS-2004B-Q-31,
GN-2004B-Q-16,
GS-2005A-Q-11,
GN-2005A-Q-11,
GS-2005B-Q-6,
GN-2005B-Q-7,
GN-2006A-Q-7,
GN-2006B-Q-10,
GN-2007A-Q-8) 
obtained at the Gemini Observatory, which is
operated by the Association of Universities for Research in Astronomy,
Inc., under a cooperative agreement with the NSF on behalf of the
Gemini partnership: the National Science Foundation (United States),
the Particle Physics and Astronomy Research Council (United Kingdom),
the National Research Council (Canada), CONICYT (Chile), the
Australian Research Council (Australia), CNPq (Brazil) and CONICET
(Argentina).
Based on observations obtained at the W.M. Keck
Observatory, which is operated as a scientific partnership among the
California Institute of Technology, the University of California and the
National Aeronautics and Space Administration. The Observatory was made
possible by the generous financial support of the W.M. Keck Foundation.
}
}

\author{T. Kronborg\inst{1,2} 
\and D.~Hardin\inst{1}
\and J.~Guy\inst{1}
\and P.~Astier\inst{1}
\and C.~Balland\inst{1,3}
\and S.~Basa\inst{4}
\and R.~G.~Carlberg\inst{5}
\and A.~Conley\inst{5}
\and D.~Fouchez\inst{6}
\and I.~M.~Hook\inst{7,8}
\and D.~A.~Howell\inst{9,10}
\and J.~J\"onsson\inst{7}
\and R.~Pain\inst{1}
\and K.~Pedersen\inst{2}
\and K.~Perrett\inst{5}
\and C.~J.~Pritchet\inst{11}
\and N.~Regnault\inst{1}
\and J.~Rich\inst{12}
\and M.~Sullivan\inst{7}
\and N.~Palanque-Delabrouille\inst{12}
\and V.~Ruhlmann-Kleider\inst{12}
} 

\titlerunning{Gravitational lensing of the SNLS sample.}

\institute{
LPNHE, Universit\'e Pierre et Marie Curie, Universit\'e Paris Diderot, CNRS-IN2P3, 4 place Jussieu, 75252 Paris Cedex 05, France 
\and Dark Cosmology Centre, Niels Bohr Institute, University of Copenhagen, Juliane Maries Vej 30, DK-2100 Copenhagen, Denmark 
\and University Paris 11, F-91405 Orsay, France 
\and LAM, CNRS, BP8, Traverse du Siphon, 13376 Marseille Cedex 12, France 
\and Department of Physics and Astronomy, University of Toronto, 50 St. George Street, Toronto, ON M5S 3H4, Canada 
\and CPPM, CNRS-IN2P3 and Universit\'e Aix-Marseille II, Case 907, 13288 Marseille Cedex 9, France 
\and Department of Astrophysics, University of Oxford, Keble Road, Oxford OX1 3RH, UK 
\and INAF -- Osservatorio Astronomico di Roma, via Frascati 33, 00040 Monteporzio (RM), Italy 
\and Las Cumbres Observatory Global Telescope Network, 6740 Cortona Dr., Suite 102, Goleta, CA 93117 
\and Department of Physics, University of California, Santa Barbara, Broida Hall, Mail Code 9530, Santa Barbara, CA 93106-9530 
\and Department of Physics and Astronomy, University of Victoria, PO Box 3055, Victoria, BC V8W 3P6, Canada 
\and IRFU, CEA/Saclay, 91191 Gif-sur-Yvette Cedex, France 
} 
\authorrunning{T.~Kronborg {\it et al.}, SNLS collaboration}
\date{Received November 09, 2009; accepted January 29, 2010}

\abstract {} { 
The observed brightness of Type Ia supernovae is affected by
gravitational lensing caused by the mass distribution along the line of
sight, which introduces an additional dispersion into the Hubble
diagram. We look for evidence of lensing in the SuperNova Legacy Survey 3-year data set.
}
{ 
We investigate the correlation between the residuals from the Hubble
diagram and the gravitational magnification based on a 
 modeling of the mass distribution of foreground galaxies. A deep photometric catalog, photometric redshifts, and well established mass luminosity relations are used.
}
{ 
We find evidence of a lensing signal with a $2.3\,\sigma$
significance.  The current result is limited by the number of SNe, their
redshift distribution, and the other sources of scatter in the Hubble diagram.
Separating the galaxy population into a red and a blue sample has a positive impact on the significance of the signal detection. 
On the other hand, increasing the depth of the galaxy catalog, the 
 precision of photometric redshifts or reducing the scatter in the mass luminosity relations
 have little effect.
  We show that for the full SuperNova Legacy Survey sample  ($\sim$400
spectroscopically confirmed Type Ia SNe and $\sim$200 photometrically identified Type Ia SNe), there is an 80\% probability of detecting the
 lensing signal with a 3 $\sigma$ significance.
}  {}

\keywords{supernovae: general -- cosmology: observations -- gravitational lensing}

\maketitle

\section{Introduction}

Type Ia supernovae (SNe~Ia) have become an essential tool of 
observational cosmology \citep{Riess98b, Perlmutter99, Astier06, Wood-Vasey07, Riess07, Freedman09, Kessler09}. By
studying the distance-redshift relation of a large number of 
SNe~Ia over a wide range in redshift, the equation of state of dark energy 
can be measured. Although SNe Ia can be calibrated to be good standard
candles, they are affected by systematic effects such as
extinction by dust and gravitational lensing. 
Because of gravitational
lensing, most supernovae will be slightly demagnified and some will be
significantly magnified due to the mass inhomogeneities along the line
of sight. 

 This causes an additional
dispersion in the observed magnitudes and thus in the Hubble diagram. The effect
is greater at high redshift, but is still less than the scatter in the Hubble diagram for
 a redshift up to 1, as we see in this paper. 

Magnification of SNe Ia can be estimated in 2 ways. 
The Hubble diagram residuals, computed assuming the best fit cosmological model, give

an indirect measure of the SN
magnification. The magnification can, on the other hand, also be estimated by

modeling the foreground galaxy mass distribution using galaxy 
photometric measurements together with derived mass-luminosity
relations for galaxies and dark matter halo models. If a correlation
between these two estimates is found, it could give interesting insight in
the modeling of the mass distribution of the foreground galaxies. This will be discussed in another paper \citep{Jonsson09}

The lensing signature in supernovae samples has already been
sought for. \cite{Williams04} 
correlated the brightness of high-z supernovae from the High-z
Supernova Search Team and the Supernova Cosmology Project with the
density of the foreground galaxies. They found that brighter
supernovae preferentially lay behind over-dense regions at a 99 \% CL.
However, they find a lensing contribution to the scatter in the Hubble
diagram of 0.11 mag (for a median redshift below 0.5) which is 
larger than expected (for example from lensing statistics), as noted by the authors themselves. 
\cite{Menard05} remark that the measured size
of the effect is also incompatible with shear variance measurements.
Finally, such a large magnification effect would cause a significant
increase of Hubble diagram scatter with redshift, undetected today.
\cite{Wang05} derived
the expected weak lensing signatures of Type Ia SNe by convolving the
intrinsic distribution in peak luminosity with expected magnification
distributions and compared the expected and measured 
skewnesses of the residual distribution of 110 high and low-z SNe 
from the Riess sample \citep{Riess04}. The signal found is dominated
by high redshift events uncorrected for the NICMOS non-linearity.
Its origin remains unknown and its significance is not provided.
 Partly using the same supernova sample, \cite{Menard05} 
searched for a correlation using SDSS photometry of the foreground
galaxies and found no evidence for a correlation. 
More recently \cite{Jonsson06} found a weak correlation with a confidence 
level of $90\%$ using high-z supernovae from the GOODS fields.  
A firm detection of this effect therefore remains to be found.

The SNLS sample is currently the
  best suited sample for a possible detection of such a signal thanks to its
large number of SNe~Ia at high redshift \citep{Jonsson08}.

In this paper, we investigate the possible correlation between the
magnification of the SNLS SNe derived from photometric measurements of
the foreground galaxies (together with chosen mass-luminosity relations)
and the residuals from the Hubble diagram.

Evaluating the expected magnification by foreground galaxies can be 
schematically split into three steps : first obtaining redshifts 
and absolute magnitudes of the foreground galaxies, then converting 
these informations into mass, and finally turning this 
set of mass and redshift estimates into a magnification estimate. 

The outline of the paper is as follows. 
$\S$\ref{sec:mass-luminosity-relations} introduces the mass-luminosity
relations used in this analysis.
In $\S$\ref{sec:data} the SNLS 3-year data set 
is presented. $\S$\ref{sec:estimating-magnifications}
describes the analysis steps to estimate the supernovae magnifications, 
while in $\S$\ref{sec:lensing-signature} we present the measured correlation 
 and compare it with expectations from simulations.
Finally, we summarize in $\S$\ref{sec:conclusion}.

\section{Mass Luminosity relations}
\label{sec:mass-luminosity-relations}

There are several ways of inferring the combined mass of a galaxy and its dark matter halo using the measured luminosity. 
We make use of two approaches in this paper.

i) The line-of-sight velocity dispersions in elliptical galaxies and rotational velocities
in spiral galaxies are correlated with their luminosities. Those are the well established Faber-Jackson \citep{Poveda1961, Fish1964, Faber1976} and Tully-Fisher \citep{Tully1977, Haynes1999} relations . 

ii) Galaxy-galaxy lensing measurements allow a relation to be determined between luminosity and mass once a functional form is assumed for the mass density profile.

We present below recent results for these approaches and compare them assuming a Singular Isothermal Sphere dark matter density profile (hereafter SIS) which has proven to be a good fit
to lensing galaxies \citep{Koopmans06}.

\subsection{Tully-Fisher relation for spiral galaxies}
\label{sec:tully-fisher}
We have chosen to use the \cite{Boehm04} results for
the Tully-Fisher (TF) relation. The maximum rotational velocities of 77
spiral galaxies in a redshift range of 
$0.1 < z < 1.0$  were measured in the FORS Deep Field. Anchoring the TF relation at low $z$ using the results from \cite{Pierce1992}, they obtain the
following relation between the maximum rotation velocity $v_{\rm max}$
and the absolute magnitude $M_B$ of the galaxy in the rest-frame $B$-band (in a Vega magnitude system):
\begin{equation}
\log v_{\rm max}=-0.134\left(M_{B}+(3.61\pm0.24)+(1.22\pm0.56)\cdot z\right)
\label{eq:tully-fisher}
\end{equation} 
The observed scatter of $M_{B}$ about this relation is $0.41$ mag (r.m.s.).
The redshift dependence expresses the fact that a galaxy of a given mass 
was brighter in the past as it hosted a younger stellar population \citep{Barden03,Milvang-Jensen03, Bamford06, chiu08}.

The maximum rotational velocity $v_{\rm max}$ of spiral galaxies, measured at large galactic radii, is dominated by the dark matter mass. Assuming a SIS dark matter density profile for which the rotational velocity is constant, we have $v_{\rm max} \simeq v_{\rm rot}(r \rightarrow \infty) = \sqrt{2} \sigma$, where $\sigma$ is the one-dimensional r.m.s. of the velocity distribution.

\subsection{Faber-Jackson relation for elliptical galaxies}
\label{sec:faber-jackson}
We adopt the Faber-Jackson relation (FJ) by \cite{Mitchell05} (hereafter M05) derived for a sample of $\sim30,000$ elliptical galaxies using
SDSS data. The selection criteria for the sample and the estimate of the
velocity dispersion are described in \cite{Bernardi03I}. The observed
velocity dispersion has been determined by analyzing the integrated
spectrum of the whole galaxy and aperture corrected to a standard
effective radius.
M05 find the following relation between the velocity dispersion and
the absolute magnitude $M_{r}$ of the galaxy in the rest-frame $r$-band  (in a AB system):
\begin{equation}
<\log(\sigma)>=2.2-0.091(M_{r}+20.79+0.85 z)
\label{eq:faber-jackson}
\end{equation}
As for the TF relation, the redshift dependence accounts for the evolution of
the age and hence luminosity of the stellar population. 
The scatter in the FJ relation induces an uncertainty in the estimate of the 
velocity dispersion which has been given by \cite{Sheth03}:
\begin{equation}
\mathrm{r.m.s.}(\log\sigma)= 0.79 (1+0.17(M_{r}+21.025+0.85z)) \nonumber
\end{equation} 
The measured velocity dispersions are aperture-corrected central velocity
 dispersions which are almost equal to the dark matter
velocity dispersions (see \citealt{Franx1993, Kochanek1994}), so we can identify those
 measurements with the velocity dispersion parameter of the SIS profile. 

We choose to translate the FJ r-band relation into a restframe B-band relation since restframe B is covered with SNLS griz bands up to $z\simeq1$.
To convert the SDSS $r$-band absolute magnitudes in the AB system to standard $B$-band
Vega absolute magnitudes, a typical color $M_{B}-M_{r}$ for
ellipticals in the AB system is estimated yielding $M_{B}-M_{r}=1.20$\footnote{With the galaxy spectral sequence described in $\S$\ref{sec:photometric-redshifts}, and for the range of rest-frame $U\!-\!V$ colors we consider to select photometrically elliptical galaxies, we obtain on average $M_{B}-M_{r}=0.86$ with an r.m.s of 0.07. This corresponds to an average change of $\sigma$ of 7\% with an r.m.s of 1\% (see Eq.~\ref{eq:faber-jackson}). We ignore this small correction in the following.}
\citep{Gunnarsson06} and an AB to Vega relation $B_{\rm AB}=B_{\rm Vega}-0.12$
is adopted.

\subsection{Galaxy-galaxy lensing mass estimates}

The galaxy-galaxy lensing signal manifests itself by images of
background (faint) galaxies being distorted by foreground (brighter)
galaxies \citep{Hoekstra04, Hoekstra05, Parker05, Kleinheinrich04, Mandelbaum06}. 
Unfortunately, one can only study ensemble averaged
properties because the weak-lensing distortion induced by an
individual galaxy is too small to be detected. The measured quantity is
the mean tangential shear which is proportional to the total lens 
galaxy-halo mass.  Since the lensing signal depends on the angular
diameter distance between observer, lens and source, 
galaxy-galaxy lensing results strongly depend on the properties of the galaxy sample, which differ for different surveys. 

We use the galaxy-galaxy lensing result from \cite{Kleinheinrich04}
(hereafter K06). The data is taken from the COMBO-17 survey which
consists of observations in five broad-band filters (UBVRI) and 12
medium-band filters. They report a mass to luminosity relation for the full
sample probed out to a maximum radius of 150~h$^{-1}$kpc when modeling
the lenses as SIS profiles: 
\begin{equation}
\sigma=156^{+18}_{-24}\left(\frac{L}{10^{10}h^{-2}L_{r\odot}}\right)^{0.28^{+0.12}_{-0.09}} \mathrm{km}.\mathrm{s}^{-1}
\label{eq:galaxy-galaxy-lensing}
\end{equation} 
where $L$ is the luminosity of the galaxy. The fiducial
luminosity, $L^{*}=10^{10}h^{-2}L_{r\odot}$ is given in the SDSS
$r$-band. For a conversion to the $B$-band, they have calculated that the
galaxies in their sample with a fiducial luminosity of
$L_{*}=10^{10}h^{-2}L_{B\odot}$ in the $B$-band have a fiducial
luminosity of $L_{*}=1.1\times10^{10}h^{-2}L_{r\odot}$ in the SDSS
$r$-band.
Note that for the SIS profile, the halo is characterized by its velocity dispersion which is the key parameter entering in the estimation of the magnification (see section \ref{sec:gravitational-magnification}). 

\subsection{Comparison assuming a Singular Isothermal Sphere profile}

\begin{figure}[h]
\begin{center}
\includegraphics[width=1.1\linewidth]{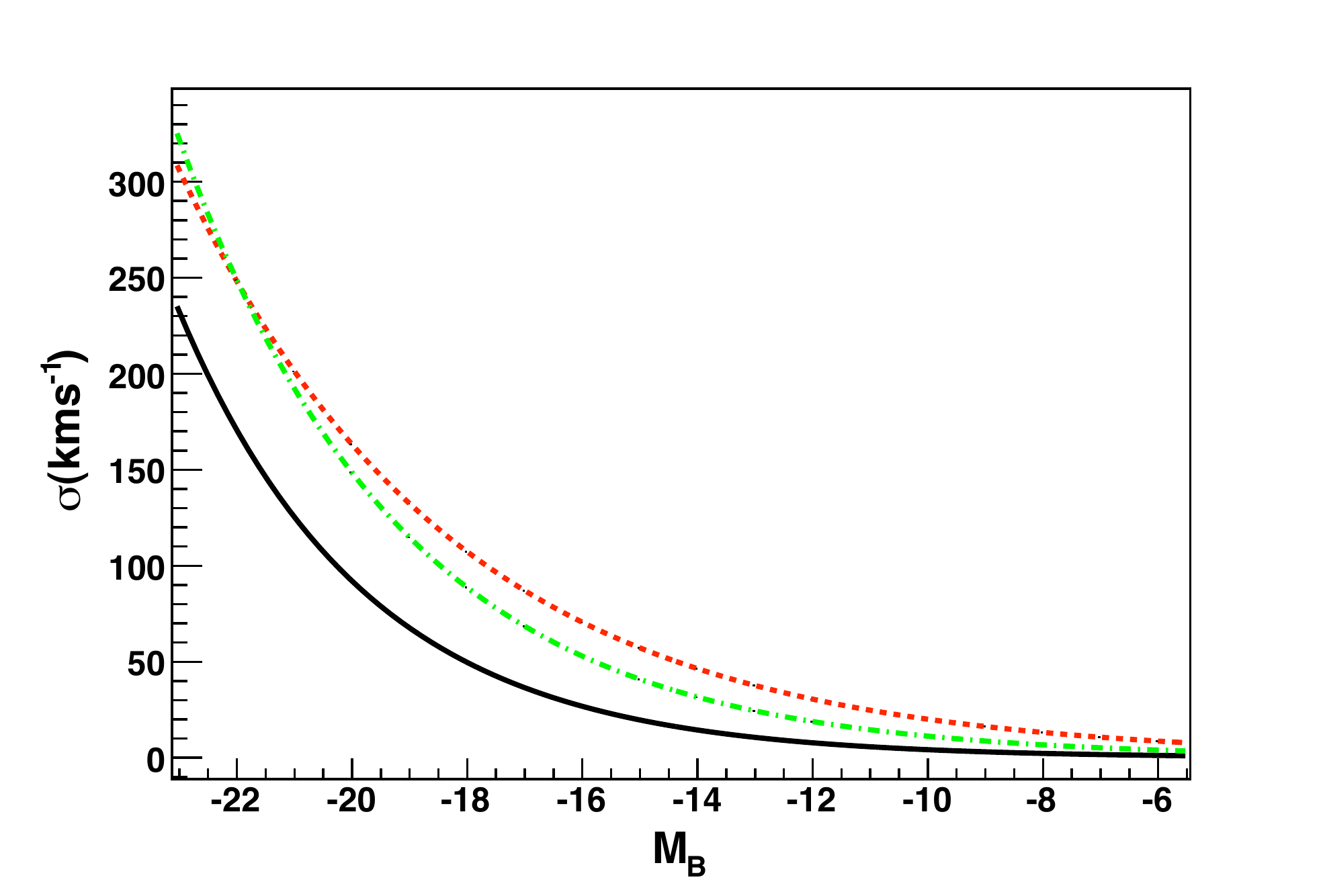}
\caption{Velocity dispersions as a function of the absolute magnitude in the $B$-band for three different mass-luminosity relations 
at a redshift of $z=0.4$. In green (dash-dotted) the K06 relation, in red (dotted) the FJ relation and in
black (solid) the TF relation. The plot shows the velocity dispersion in km s$^{-1}$
as a function of absolute magnitude in the $B$-band. }
\label{lum_mass}
\end{center}
\end{figure}

In Figure \ref{lum_mass}, we show the velocity dispersion as a
function of absolute $B$-band magnitude for the three different
mass-luminosity relations
(Eq.~\ref{eq:tully-fisher},\ref{eq:faber-jackson},\ref{eq:galaxy-galaxy-lensing}
for TF, FJ relations and galaxy-galaxy lensing results of K06) assuming a SIS
dark matter mass profile. The mean redshift of the lens galaxies in
the K06-relation is $\sim0.4$. For a given $B$-band luminosity,
elliptical galaxies are more massive than spirals so the FJ relation
 naturally gives higher velocity dispersion than
the TF relation. The K06 relation and the FJ relation appear quite similar.
 For high luminosity galaxies
the difference in mass estimate from the three relations can lead to very different magnifications. As an example, for a bright galaxy
of absolute magnitude M$_{B}$=-22, the velocity
dispersions range from 170~km s$^{-1}$ (TF) to 249~km s$^{-1}$ (FJ) and 248~km s$^{-1}$
(K06).
As a consequence, in order to obtain accurate estimates of the magnification it is important to take into account the galaxy type.  \\ 

\section{Data}
\label{sec:data}
The SuperNova Legacy Survey consists of an imaging survey which is
part of the (deep component of the) Canada-France-Hawaii Telescope
Legacy Survey\footnote{see http://www.cfht.hawaii.edu/Science/CFHLS} and a spectroscopic survey done on 8m class telescopes. The imaging was done with the MegaCam imager (360
Megapixels, 1~deg$^{2}$) which detects and monitors the light curves of
the SNe, over four different fields of 1~deg$^{2}$ (called D1, D2, D3, D4).
D1 overlaps partly with the VVDS survey\footnote{http://www.oamp.fr/virmos/vvds.htm}, D2 with the COSMOS survey\footnote{http://cosmos.astro.caltech.edu/}, and
D3 with the DEEP2 survey\footnote{http://deep.berkeley.edu/}. We will later
use the spectroscopic galaxy redshifts from the DEEP2 and VVDS surveys
to establish and test photometric galaxy redshifts.

The "rolling" search method was used for the $griz$ bands, 
which means observing the same field every 3-4 days during dark and gray time
(typically for a period of $\sim16$ days around new moon)
for as long as it remains visible (from 5 to 7 months a year). Images in $u$ are also used in this analysis. 
The images are used both to monitor the SN light curves
and to build photometric catalogs of the objects in the fields, in
particular the galaxies.

For each field, the total exposure times used
to build the galaxy catalogs for this analysis amount to more than 20 h
per band, reaching typically 60 hours in the $i$-band.

Spectroscopy of the supernovae is 
crucial in order to obtain
redshifts, and to confirm the type of each supernova. The
spectroscopic follow-up for the SNLS was done with the VLT~\citep{Baumont08,Balland09}, Gemini~\citep{Howell05,Bronder08} and
Keck telescopes~\citep{Ellis08}.
The 3-year data set used in this paper consists of 233
spectroscopically confirmed Type Ia supernovae used for cosmological
analyses. A detailed description of the SN survey
and the SN data analysis methods is provided in \cite{Astier06,Guy09}.
An important feature, relevant here, is that the
SNLS Hubble diagram exhibits a r.m.s scatter of 0.16 mag.

\section{Estimating the supernovae magnification}
\label{sec:estimating-magnifications}

We describe in this section the analysis steps followed to calculate the supernovae magnifications.
We  need a catalog of the field
galaxies  with an estimate of the redshift and the rest-frame $B$, $V$ and $U$
band absolute magnitudes for each galaxy.  Computing the restframe magnitudes requires, in addition to the redshift, 
the knowledge of the galaxy SED.
 The rest-frame $B$-band absolute
magnitude is used to convert the luminosity of the galaxy into a
mass estimate using the mass-luminosity relations introduced in
section \ref{sec:mass-luminosity-relations} whereas the $U\!-\!V$ color is
 used to separate the galaxies into a red and a blue population. 
The photometric catalog is described in $\S$\ref{sec:catalog}. The method we have used to determine photometric redshifts
and absolute magnitudes is accounted for in $\S$\ref{sec:photometric-redshifts}, the galaxy photometric classification is presented in $\S$\ref{sec:spiral-elliptical-classification}, and the estimation of the SN magnification is done in 
$\S$\ref{sec:gravitational-magnification}.

\subsection{Galaxy photometric catalog}
\label{sec:catalog}

The galaxy catalogs are built on deep image stacks in the $ugriz$
 filters. The deep stacks are constructed by selecting 80 \% of
the best quality images (6241 images for the four fields). Transmission
and seeing cuts (e.g. {\sc fwhm} $< 1.15\arcsec$) are applied. Because we
have fewer exposures in the $u$-band than in the others, less
stringent quality cuts are applied to these images. The selected
images are co-added using the {\sc swarp v2.10}
package\footnote{http://terapix.iap.fr/soft/swarp/}. The source
detection and photometry is performed using SExtractor V2.4.4 (\citep{Sex}) in
double image mode. The detection is made in the i band.

 A cut on the signal-to-noise ratio 
in the i-band, $S/N>15$ has been made so as to maximize the signal detection while
minimizing the many spurious detections around stars halos. We then
use the AUTO SExtractor flux, computed in an elliptic aperture to
extract the galaxies.
The different cuts lead to a limiting magnitude
$i=25$ (Vega magnitude system).
Zero points are computed using circular photometry on a tertiary stars
catalog, described in  \citep{Regnault09}.

 Two categories of objects
have to be identified in
 the catalog : stars, and the host galaxies of the SNe.
In order to identify stars, we estimate the second moments of all objects in the catalog using a 2D Gaussian fit. We then look for an accumulation
in the second moment $m_{xx}-m_{yy}$ space, taking into account the
variations of the PSF across the focal plane. Objects in this clump
are assumed to be stars.

Since the photo-z of the host galaxy is less precise than the spectroscopic 
z of the SN it is possible to have the configuration of the host galaxy being 
in the line-of-sight of the supernova and hence wrongly contributing to its magnification. 
As a consequence it is important to identify and exclude the host galaxy from the analysis.
Note that identifying the host galaxy on the images is not always obvious.
The host galaxy is identified using two criteria: the minimum distance to the
 supernova location and the match between the photometric redshift of the galaxy and 
 the spectroscopic redshift of the supernova.

To estimate these quantities  it is necessary to use images uncontaminated by the supernova light. 
For a given supernova, we exclude the images
taken during the same season, i.e. the 6 consecutive months during
which the field was observed. The photometric redshift is derived using the
 SNLS photometric redshift code (see section \ref{sec:photometric-redshifts}).
 A normalized distance $d$ is
computed using SExtractor shape parameters, so that $d<1$ defines the
photometric ellipse of the galaxy. When no object is found within
$d<1.3$ of the SN, the host of the SN is undetected. 
When more than one object is detected close to the
supernova location, we check the match between the galaxies
photometric and the supernova spectroscopic redshifts.
 Ambiguous cases
are flagged as problematic and can lead to exclusion of the SN if the
uncertainty in the determination of the host galaxy has an important impact
on the magnification of the SN in question. Four SNe from the sample
were initially excluded in this way.

\begin{figure*}[!t]
\begin{minipage}[t]{0.47\textwidth}

\subfigure[CFHT image.]{\includegraphics[width=0.95\linewidth]{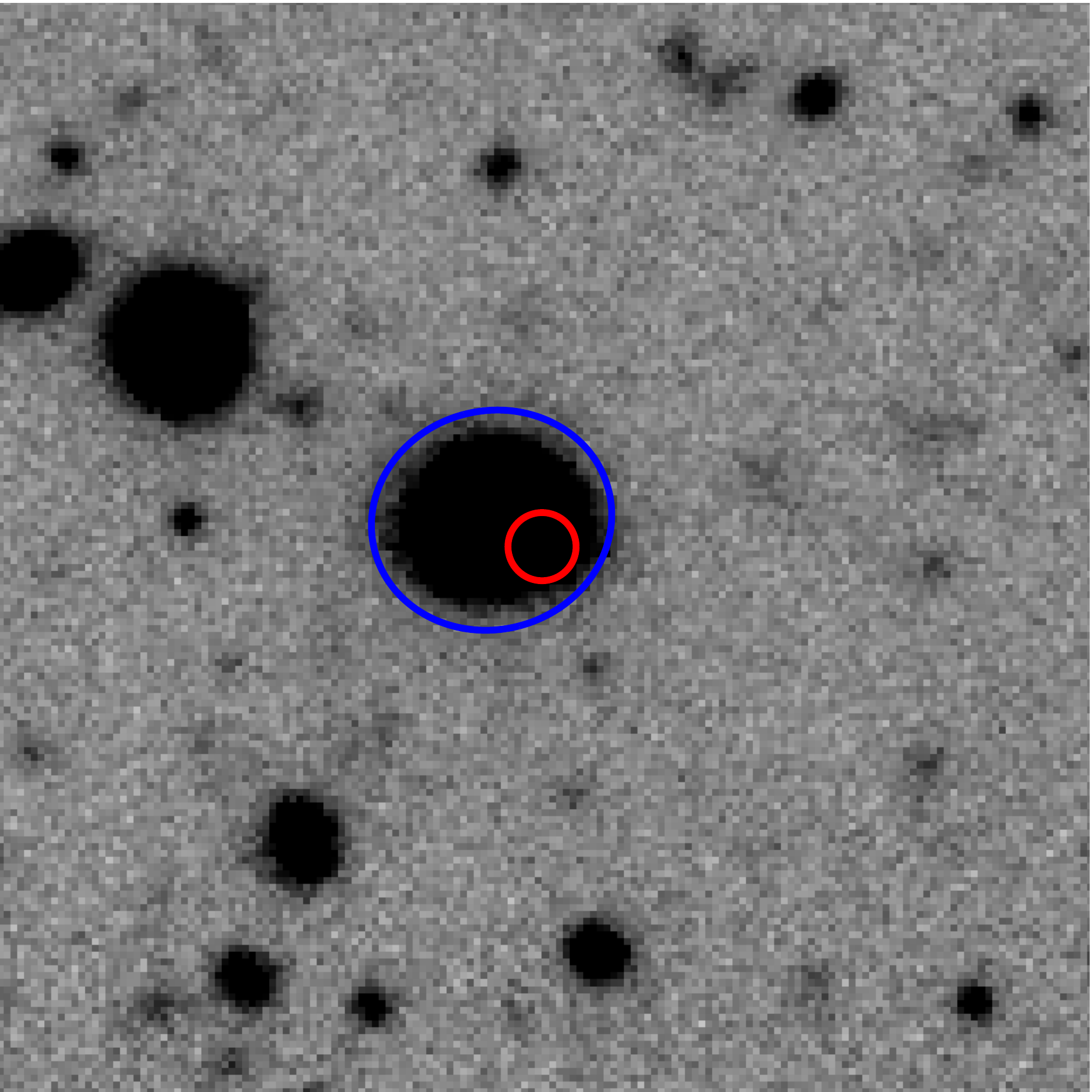}}

\subfigure[HST image from the COSMOS survey]{\includegraphics[width=0.95\linewidth]{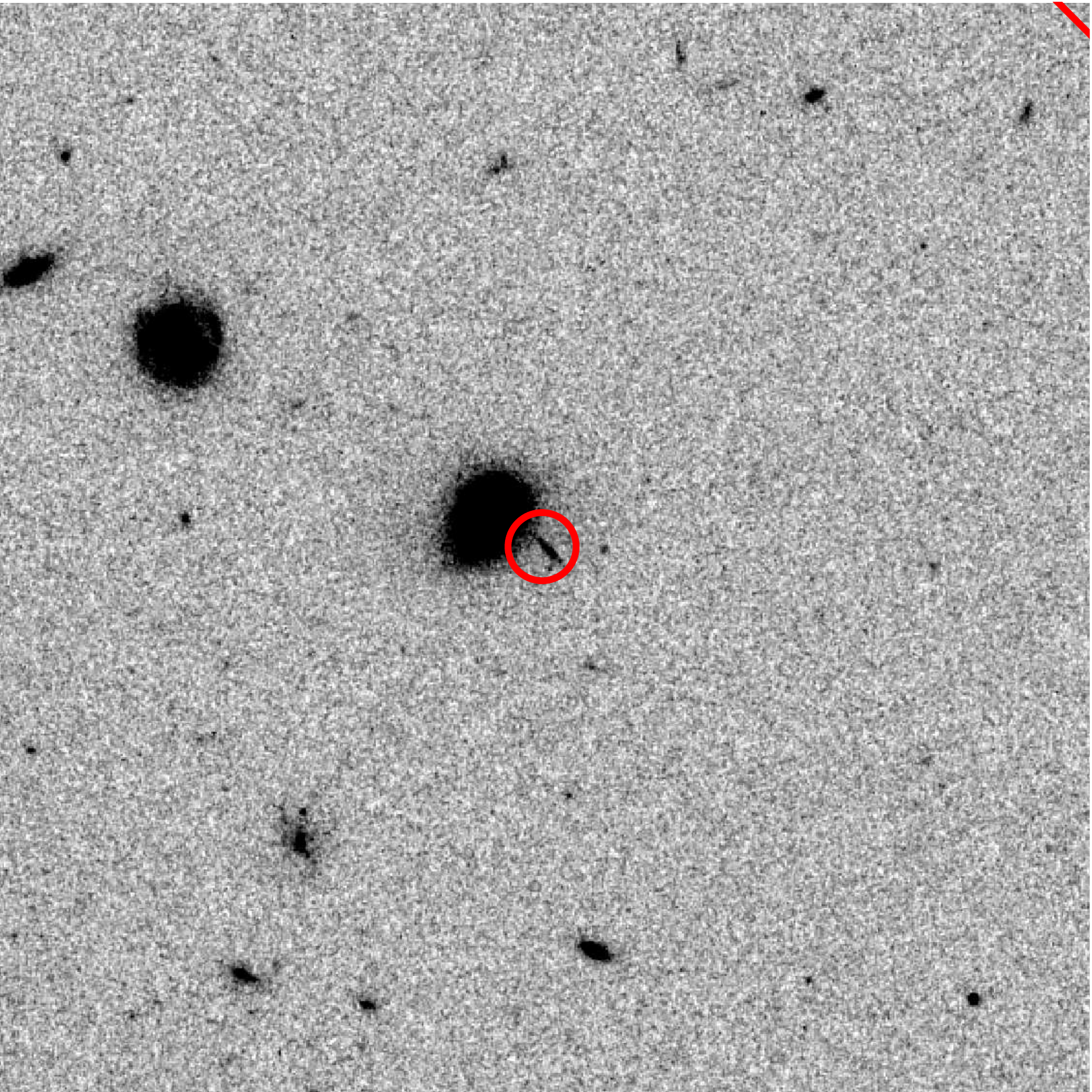}}
\caption{Images of the field around the supernova SNLS-04D2kr at $z=0.744$. The images span 30\arcsec and are oriented north-up/east-left.  A red circle indicates the supernova position. A blue ellipse indicates the closest galaxy, as detected on the CFHT image.}
\label{04D2kr}
\end{minipage}\hfill
\begin{minipage}[t]{0.47\textwidth}

\subfigure[CFHT image]{\includegraphics[width=0.95\linewidth]{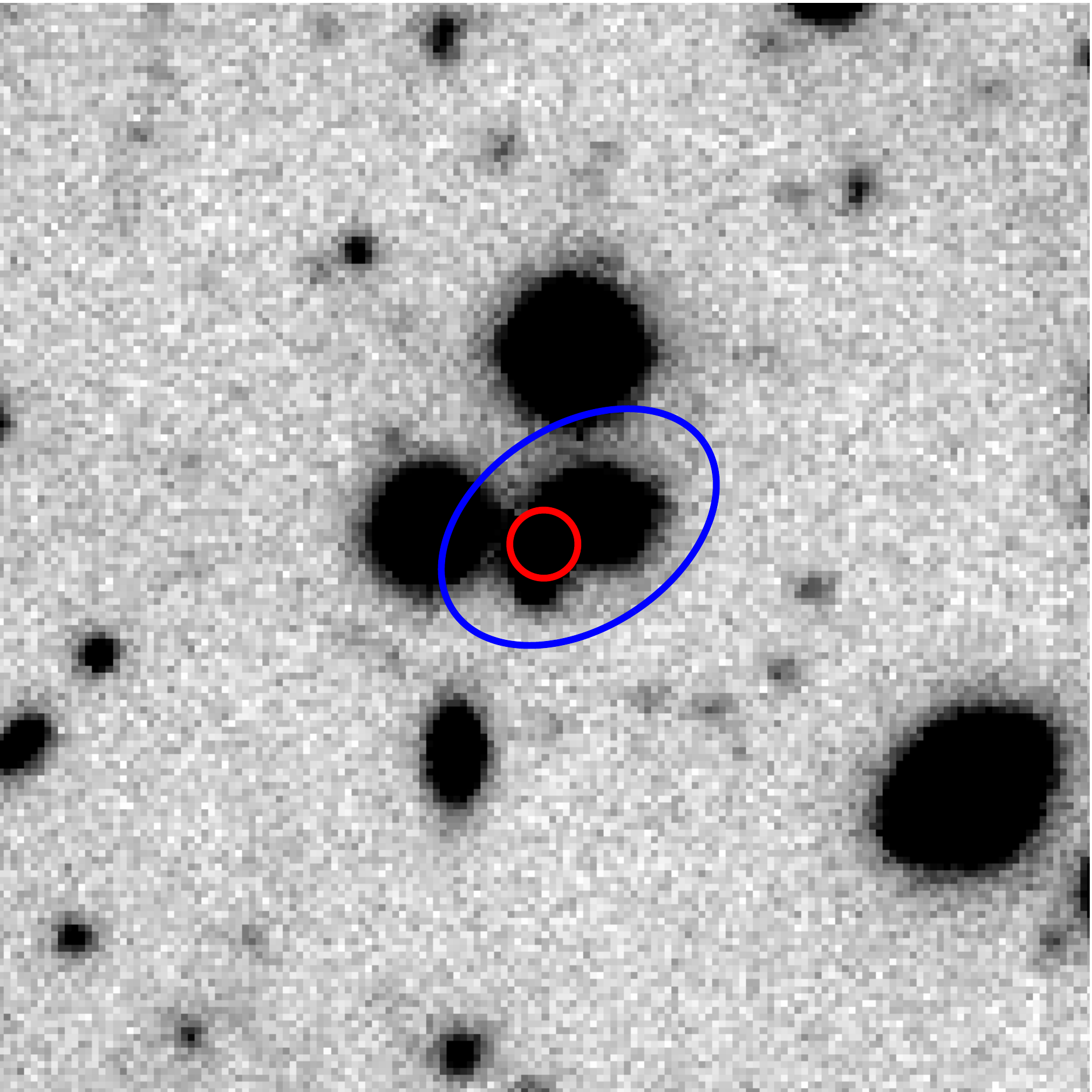}}

\subfigure[HST image from the COSMOS survey]{\includegraphics[width=0.95\linewidth]{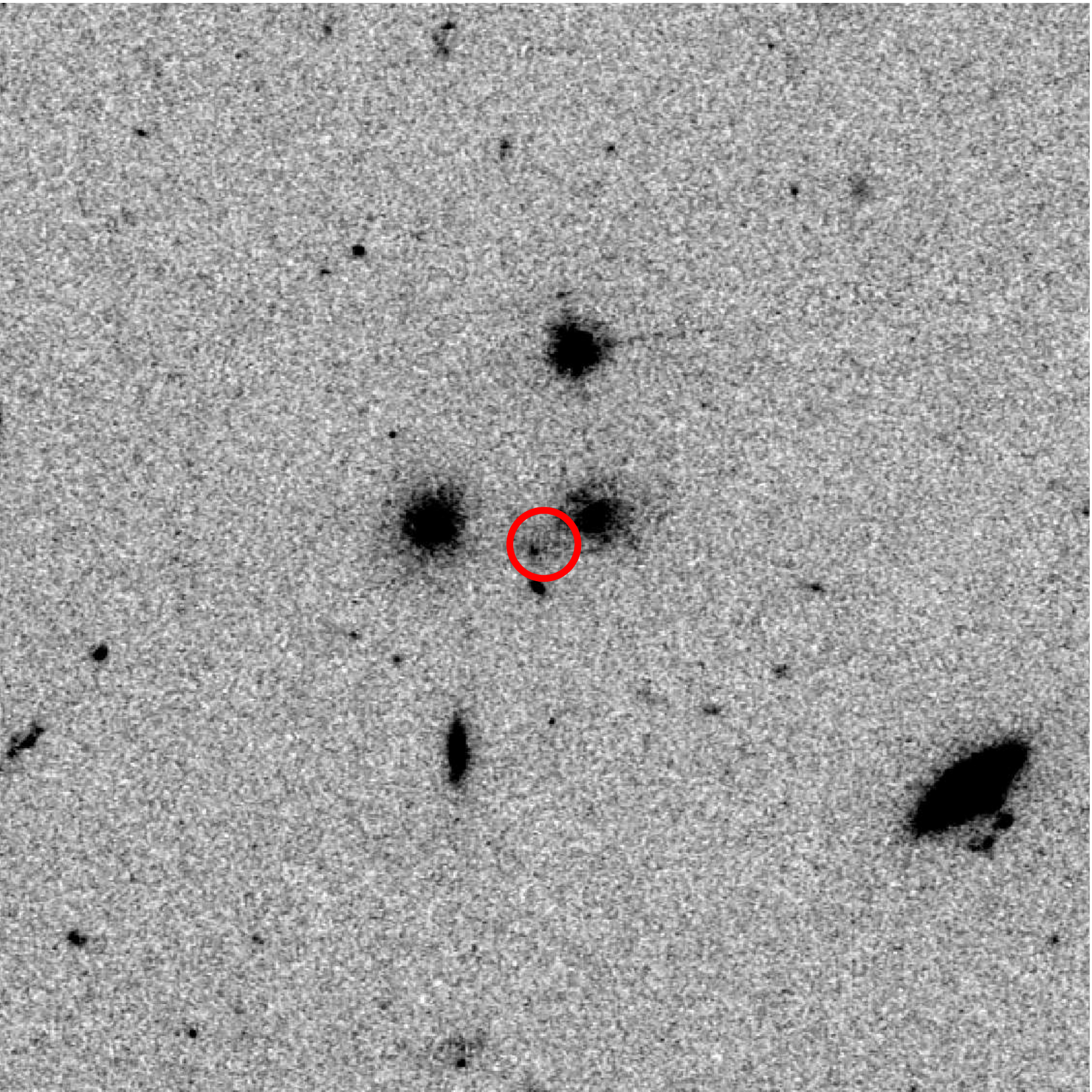}}
\caption{ Field around SNLS-05D2bt at $z=0.68$. As for SNLS-04D2kr, a smaller and fainter galaxy is conspicuous on the SN location on the HST image.}
\label{05D2bt}
\end{minipage}
\end{figure*}

However, we are fortunate to have HST imaging from the COSMOS
 field \citep{Scoville07,Koekemoer07} together with newly published high resolution redshifts \citep{Ilbert09}, also from the COSMOS for 
 galaxies in the D2 field.  As a consequence, two of the initially excluded 
 SNe have been reexamined, SNLS-04D2kr and SNLS-05D2bt. These two 
 SNe are detected very close to a presumed host galaxy
which photometric redshift does not match the spectroscopic
redshift of the SN.  
Figure \ref{04D2kr} and \ref{05D2bt} show the SN location on 
the CFHT and the HST images for the two SNe in question.

For SNLS-04D2kr at $z=0.744$, a smaller and hardly visible galaxy is
detected at the location of the SN in the HST image, very close to the
presumed -- large -- host galaxy. The redshift assigned to this large galaxy from \citep{Ilbert06}, \citep{Ilbert09}, and the SNLS photometric redshift code is $z = 0.168$, 0.228 and
0.3 respectively,  implying that the large galaxy in question is not the
host galaxy, but a foreground galaxy. The host galaxy is likely to be the
small elongated galaxy detected in the HST image.

The same situation arises for  SNLS-05D2bt at\\
 $z=0.68$ : the presumed host galaxy has a photometric  redshift of $z=0.31$ 
and $0.32$ from the SNLS photometric redshift code and \citep{Ilbert06}
respectively. 

We obtain in this way a photometric catalog in the $ugriz$ bands,
where stars and supernovae host galaxies are identified. The host
galaxy magnitudes are free from supernova light.

\subsubsection{Masks}

Halos and diffraction spikes around bright stars generate spurious galaxy detections in the catalog.
The size of the halos and the shape of the spikes are fixed by the geometry of the telescope optics,
but their intensity in the images is directly proportional to the brightness of the stars.
Also, for the brightest stars, pixels reaching their saturation level induce bleedings in the
CCD images which have to be accounted for.

\begin{figure}[htp]
\begin{center}
\includegraphics[width=0.95\linewidth]{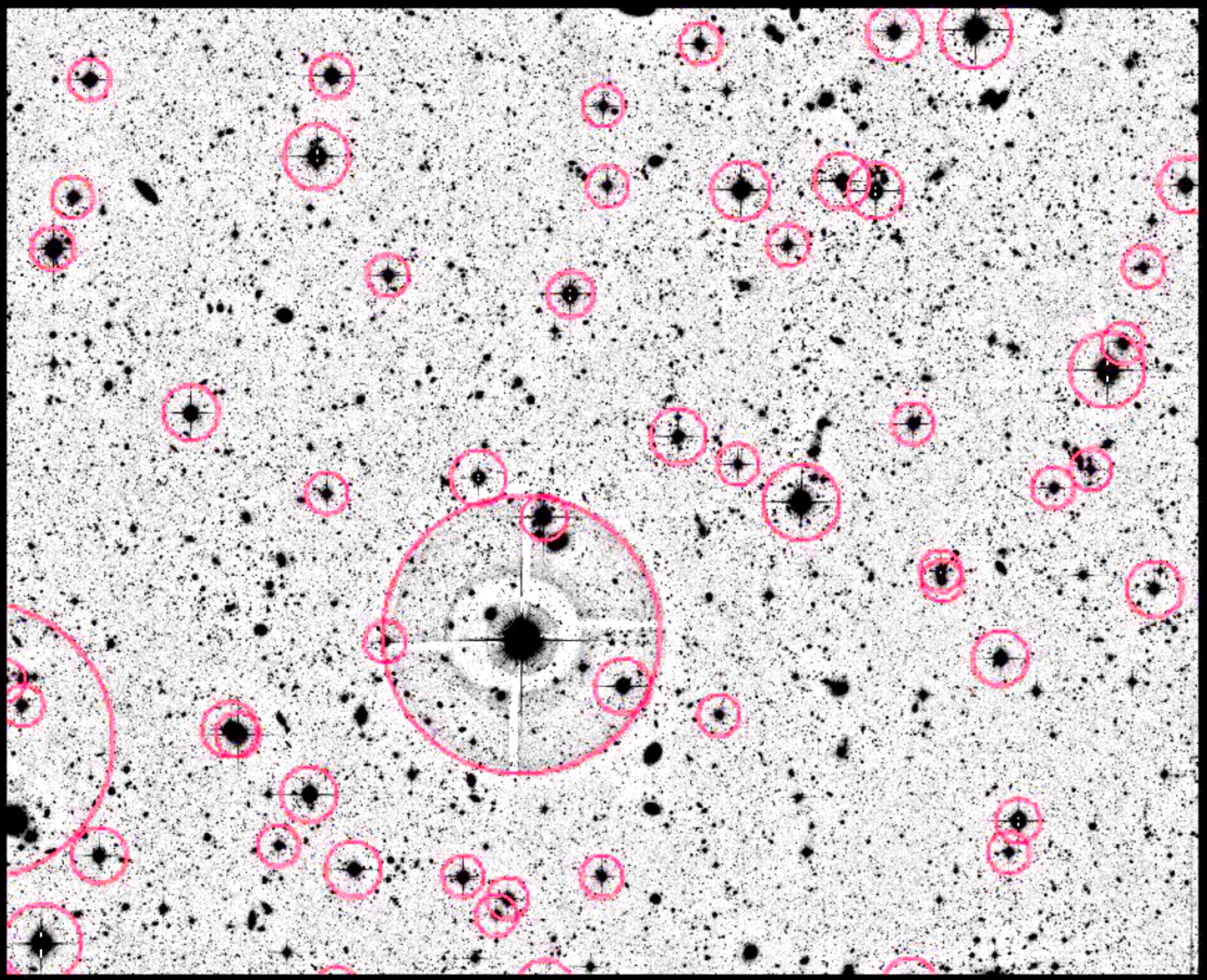}
\caption{A section of the D1 field (5000x4500 pixels or 15.4x13.9\arcmin) with the masked areas shown as purple circles.}
\label{maskfig}
\end{center}
\end{figure}

Due to those effects, circles with radius varying from 50 to 600 pixels (10 to 120\arcsec) 
 centered around bright stars are masked out. Those masked regions represent 22\%  of the field of view.
An example is displayed in Figure \ref{maskfig} for the D1 field.

For each supernova we have included galaxies within a radius of 60\arcsec
 (see $\S$\ref{sec:gravitational-magnification}). This will be referred to
as the selected area. In theory, one should exclude all
SNe where the selected area overlap a masked area, but this means
rejecting about half of the SNe. However, the effect on the magnification of small
 spurious galaxies or diffraction spikes in the outskirts of the selected area 
 is small and we chose to keep all SNe with an angular separation larger than 40\arcsec of masked areas. 
 This leads to the exclusion of 53 SNe. 

All SNe have also been looked at by eye to see
whether there are other effects that could lead to exclusion,
e.g. non-masked stars very close to the line of sight leading to the
possibility of excluding an important galaxy hidden behind the star or
leading to possible bad photometry for the surrounding
galaxies. Looking at the SNe we exclude 9 of them and as a conclusion
we keep 171 SNe out of 233 in the initial
sample.

\subsection{Photometric redshifts}
\label{sec:photometric-redshifts}

High quality photometric redshifts have been published by
\citet{Ilbert06} for the galaxies in the SNLS fields down to
$i_{\rm AB}\simeq24$. In these catalogs, fluxes of galaxies hosting 
supernovae are affected by the supernovae light, leading sometimes to unreliable
typing and redshift for these galaxies.  We also need the SED  used to derive the photometric redshift, so as to compute
the restframe magnitudes. Finally, we  need to be able to propagate easily the
uncertainties of the photometric measurements to the photo-z and the
absolute magnitude estimations.  For these reasons, and so as to
control the error propagation path, we have chosen to derive the
photometric redshifts and the absolute magnitudes as follows.

We first define a continuous one parameter ($a_{*}$) galaxy spectral
sequence, $F(a_{*}, \lambda)$.  Indeed as we have two other parameters
to estimate (magnitude and redshift), we cannot afford more than a
single parameter to index the diversity of galaxies in order to keep
one constraint, when, even though 5 bands are used, only 4 are well measured ($griz$, $u$ being
significantly shallower)~\footnote{It would be conceivable to add an extinction
parameter if we restricted our scope to bright galaxies.}. To define
this spectral template, we use the galaxy evolution 
model PEGASE.2~\citep{Fioc99}, using a variety of 
 galaxy SFR law $\propto (t/\tau)\times \exp(-t/\tau)$ and galaxy ages. The  initially zero  gas metallicity
evolves through the successive generations of stars and ranges  from $Z\!\sim\!0.$ to $0.03$. 
Extinction is computed using a transfer model for an inclination-averaged disk distribution, 
 the optical depth being estimated from the mass of gas and the metallicity. We left aside the possibility to add an additional extinction.

The next step consists in optimizing the spectral sequence so as to
reproduce the observed colors of our data in the best way. 

It can be corrected to about 30 \%, so that the final template spectra do differ from the initial computed  template :  our technique is thus marginally 
sensitive to the input model.
The
training set comprises a sample of galaxies with known spectroscopic
redshift from the DEEP2 survey \citep{Davis2003, Davis2007}. The spectral
sequence consists in the galaxy flux as a function of wavelength and
the age index, and we optimize a multiplicative correction
(a smooth function of wavelength and age index), along with offsets to the
photometric zero points (which can be interpreted as differential
aperture corrections). We find a residual scatter of 0.097, 0.015,
0.035, 0.025, 0.050 mag for $ugriz$ bands respectively.

\begin{figure}[h]
\begin{center}
\includegraphics[angle=90.,width=1\linewidth]{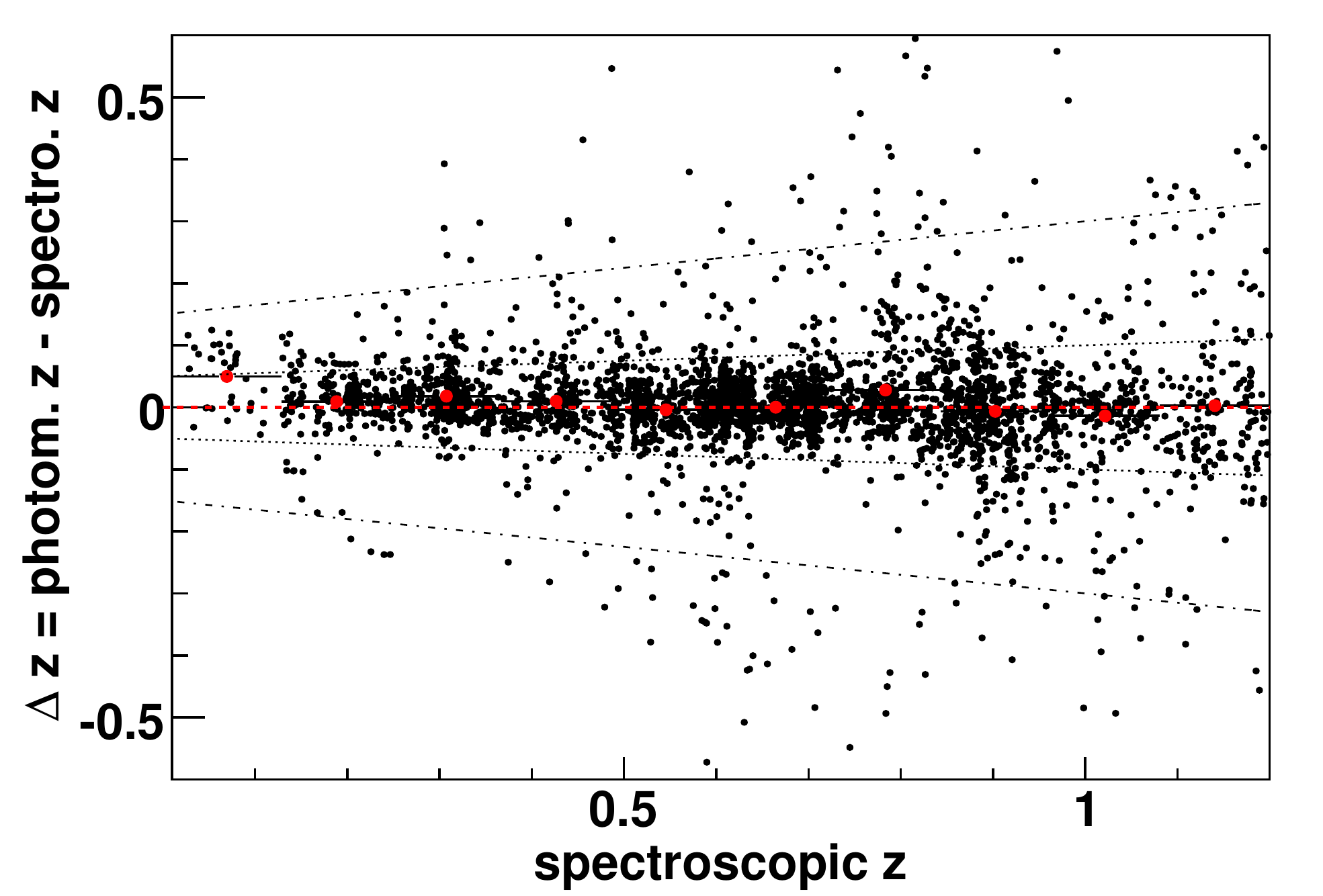}
\caption{Performance of the photometric redshift determination: the difference $\Delta z$ between spectroscopic and photometric redshift 
as a function of spectroscopic redshift. The spectroscopic
redshifts were obtained from the VVDS survey \citep{Lefevre2004}.}
\label{zspec}
\end{center}
\end{figure}

The performance of the photometric redshift computation is evaluated
using VVDS spectroscopic redshifts available for the D1 field
(3595~galaxies at $0.01<z<1.5$)~\citep{Lefevre2004}. The redshift
residuals (i.e. $\Delta$z=photometric redshift - spectroscopic
redshift) as a function of spectroscopic redshift are shown in
Figure \ref{zspec}.  For $i_{\rm AB}<24$, the fraction of catastrophic failures for
which $\Delta z/(1+z)>0.15$ is $6\%$. Eliminating
catastrophic failures, we obtain $\sigma_{\Delta z} = 0.065 $ and $\sigma_{\Delta z/(1+z)} = 0.037$. 
Note that we do not apply any prior on the photometric redshift, 

 as a large fraction
 of the SNLS galaxies are much fainter than those of the training set.

\begin{figure}[h]
\begin{center}
\includegraphics[angle=90.,width=1.0\linewidth]{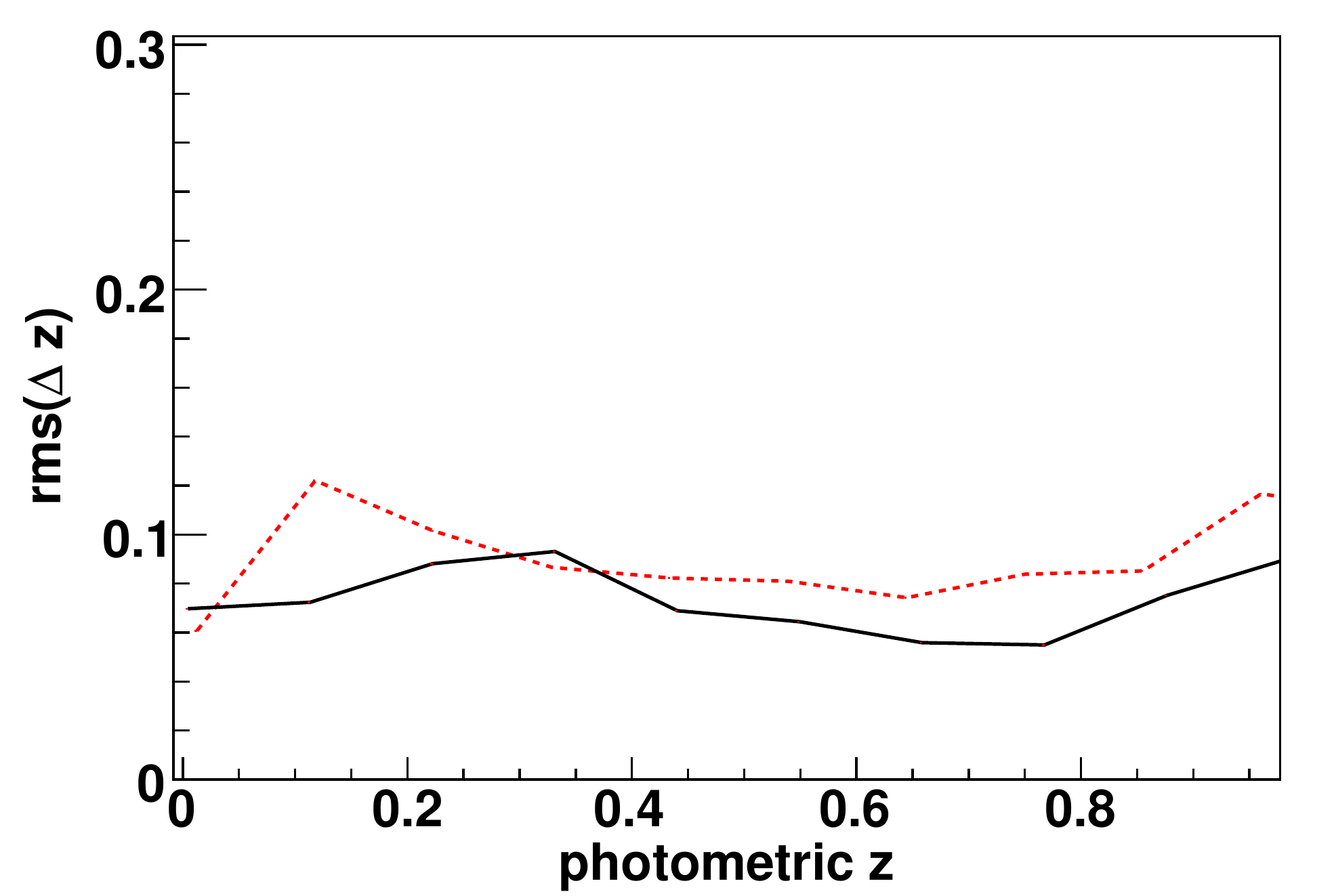}

\caption{Performance of our photometric redshift determination: the uncertainty in the estimated photo-z using the 
spectroscopic VVDS redshift as a function of photo-z (in dotted red)
and the photo-z uncertainty (solid black line)}
\label{rms_vs_z}
\end{center}
\end{figure}

Uncertainties are estimated using a Monte Carlo propagation of
magnitude uncertainties accounting for the measurement uncertainties
and the residual scatter obtained from the training procedure.  In
Figure~\ref{rms_vs_z} we compare the scatter of $\Delta z$ for
the VVDS test sample to the uncertainty on the photometric redshift derived from the
Monte Carlo propagation. Both are in reasonable agreement which
validates our method for propagating uncertainties.

For some of the galaxies it has been possible to obtain spectroscopic
redshifts from VVDS, DEEP-2 and the SNLS spectroscopic program on VLT (some field galaxies were targeted at the same time as the SN primary target during observations with FORS-2 in multi-slit mode). The D2 field overlaps with the COSMOS field and it has thus been possible to use high resolution
photometric redshifts for a large fraction of the galaxies in the D2
field \citep{Ilbert09}.

\subsection{Classification of spiral and elliptical galaxies based on their colors}
\label{sec:spiral-elliptical-classification}
The Tully-Fisher and Faber-Jackson relations are derived for spiral
galaxies and elliptical galaxies respectively and as a consequence it
is necessary to separate the SNLS galaxies into spirals and
ellipticals. Note that this classification is not strictly necessary:
we could blindly apply some sort of average mass-luminosity relation
to all galaxies. There is indeed no typing involved in
our analysis based on galaxy-galaxy lensing mass estimates.
 Separating in broad types is not more than 
a way to a priori improve the significance of a potential detection.

\begin{figure}[h]
\begin{center}
\includegraphics[angle=90.,width=1.0\linewidth]{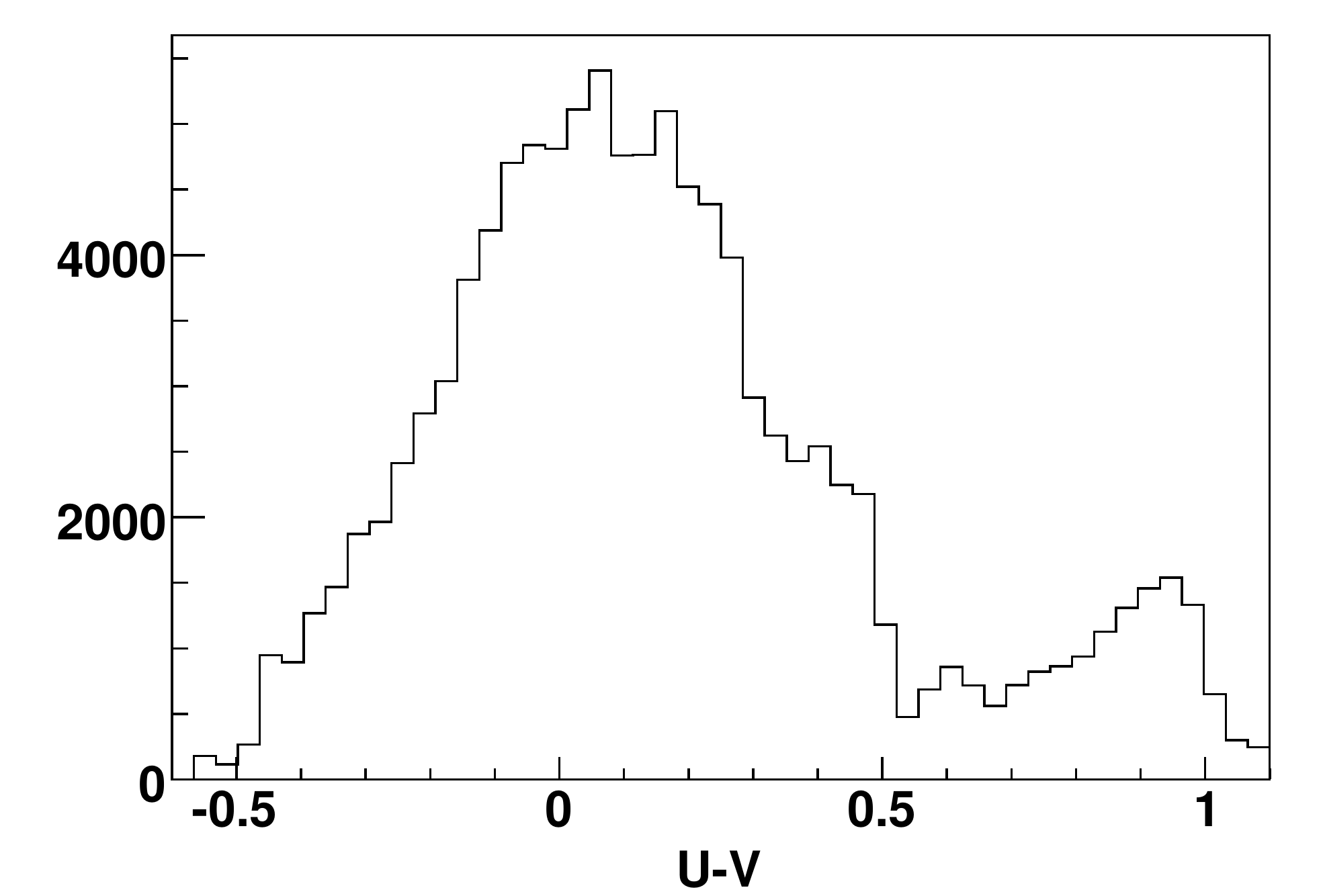}
\caption{Distribution of the rest-frame $U\!-\!V$ color for the SNLS galaxies. A cut at $U\!-\!V=0.54$ separates the distribution into red (elliptical) and blue (spiral) galaxies. }
\label{UV}
\end{center}
\end{figure}

The SNLS imaging data does not permit a good morphological classification and 
we resorted to using a simple color cut to classify our galaxies.
The rest-frame color $U\!-\!V$ is computed giving rise to two well
separated distributions (the red and the blue population). In Figure
\ref{UV} the rest-frame color $U\!-\!V$ for the SNLS galaxies is shown.
 For $U\!-\!V > 0.54$, the
galaxy is classified as an elliptical galaxy or else it is classified
as a spiral galaxy. 
Using the rest-frame $U\!-\!V$ color as a proxy for galaxy classification is obviously not optimal, 
 and the effect of miss-classifying the galaxies
 will degrade the lensing signal (see $\S$\ref{sec:correlation}).

 To evaluate whether this cut provides a good estimate for galaxy typing we compare our galaxy types to those
published in \cite{Ilbert06} and \cite{Ilbert09}. They correspond to the galaxy template type, as fitted with eventually some additional extinction, using respectively 5 and 30 band magnitude measurements. For $i_{\rm AB}\leqslant24$, we have classified $89.5\%$ of the COSMOS and $95.3\%$ of the CFHTLS elliptical galaxies as ellipticals. However $56.8\%$ and $26.9\%$ of our elliptical galaxies are classified as spiral galaxies in the COSMOS and the CFHTLS respectively, with an additional extinction of $<E(B\!-\!V)> \simeq 0.2$.
In conclusion, we identify mainly all elliptical galaxies,  but there is a contamination of our color-selected elliptical sample with red extinct spiral galaxies. 
 This contamination will dilute the lensing signal : the red spirals misidentified as ellipticals will be attributed 
  too  high a mass, falsely increasing the supernova expected gravitational magnification, and thus decreasing
   its correlation with Hubble diagram residuals.

\subsection{Gravitational magnification}
\label{sec:gravitational-magnification}
The density profile of a SIS can be written
\begin{equation}
\rho_{\rm SIS}(r)=\frac{\sigma^{2}}{2\pi r^{2}}
\end{equation}
The total mass of the halo, $m(r)=2\sigma^{2}r$, diverges and as a consequence
 we can use a truncation radius, $r_{t}$ to be able to estimate the total mass of the halo. The truncation radius is chosen to be $r_{200}=\sqrt{2}\sigma/10\,H(z)$ which is defined as the radius within which the mean mass density is 200 times the critical density.
 In this case, the convergence is given by
 \begin{equation}
 \kappa_{\rm SIS}(\theta)=\frac{\theta_{E}}{\pi\theta }\arctan \sqrt{\frac{r_{200}^{2}}{\theta^{2}D_{l}^{2}}-1}
 \end{equation}
in the thin lens approximation, where $\theta$ 
is the transverse angle of the image to the lens, and $\theta_{E}=4\pi (\sigma/c)^{2} D_{ls}/D_{s}$ is the Einstein radius. $D_{ls}$, 
$D_{s}$ and $D_{l}$ are respectively the angular distances from the lens to
the source, from the observer to the source and from the observer to the lens, and $\sigma$ is the velocity dispersion.
 In order to compute the magnification due to several galaxy lenses, 
 we use the publicly available software Q-LET~\citep{Gunnarsson04},
 which uses the multiple lens plane method.

\begin{figure}[h]
\begin{center}
\includegraphics[angle=90,width=\linewidth]{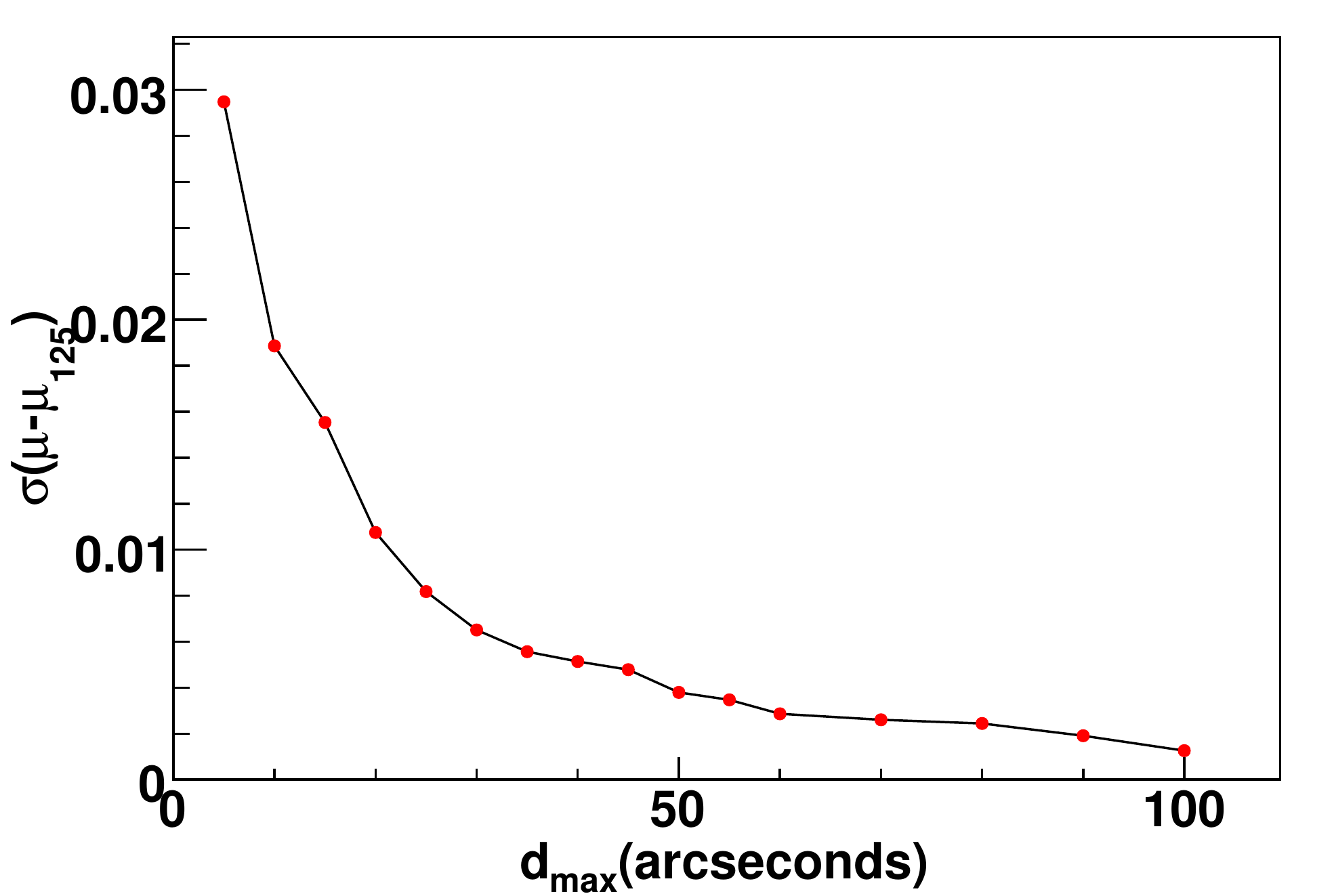}
\caption{r.m.s of the difference of magnifications $\mu(d_{\rm max})-\mu(d_{\rm max}=125\arcsec)$ as a function of $d_{\rm max}$. This simulation (from real lines of sight) is done for a source at a redshift of 1.}
\label{fig:magnification_rms_vs_r}
\end{center}
\end{figure}

We have investigated the effect on the estimated magnification of
truncating the galaxy list around lines of sight. We have considered
for this purpose random lines of sight in the galaxy catalog, and
computed for each of them the magnification for a SN at $z=1$,
considering sequentially an increasing number of galaxies in the
foreground according to their (transverse) angle $\theta$ to the SN.
Figure~\ref{fig:magnification_rms_vs_r} displays the r.m.s of the
difference of magnifications
$\mu(d_{\rm max}<125\arcsec)-\mu(d_{\rm max}=125\arcsec)$ as a function of
$d_{\rm max}$, where $d_{\rm max}$ is the maximum transverse angle of galaxies
used to compute the magnification\footnote{The maximum value of
$125\arcsec$ was limited by the maximum number of lenses that Q-LET
could handle, this has however essentially no impact on this 
analysis: the brightest 
galaxy of our sample placed at $125\arcsec$ of a supernova line of sight
causes a magnification $\mu_m < 2\ 10^{-5}$.}.
For $d_{\rm max}=60\arcsec$, the average error on the estimated
magnification due to this galaxy selection is of 0.003. This number is
much smaller than the other sources of uncertainties so that we can
safely ignore galaxies at larger angles.

\subsubsection{Normalization of the magnification distribution}

To estimate the magnification using Q-LET, the lensing galaxies have been put on 
top of a homogeneously distributed universe leading to computed gravitational
magnifications always greater than 1 compared to such a universe.
As a consequence the magnifications should be normalized in order to force the
average magnification to be equal to 1, as imposed by flux 
conservation.

For this purpose, we have considered numerous random lines of sight
 (randomly chosen source positions) 
using the true galaxy catalog for a range of source redshifts and computed 
the magnification for each of them.
The average magnification for each redshift bin (of 0.1) is recorded and subsequently used to normalize 
 the magnification estimates. 
This correction function is calculated for
each field and each mass-luminosity relation.

\subsubsection{Magnification uncertainty calculation}
\label{sec:mag_uncertainty_calculation}
The uncertainties on the magnifications are evaluated with a Monte Carlo 
propagation of the magnitude uncertainties of the galaxies from the catalog and the scatter in the mass-luminosity relations.

Magnitude uncertainties include  both measurement uncertainties and the residual scatter of the photometric redshift training (see $\S$\ref{sec:photometric-redshifts}).

A fit of the photometric redshift is performed for each Monte Carlo realization of each galaxy, except for those which have much more precise spectroscopic or photometric redshifts from other sources.
 For instance, high resolution photometric redshifts are available for galaxies of the COSMOS catalog \citep{Ilbert09}. For those, instead of fitting for the redshift, we considered random realizations of the redshift within the errors given in their paper, and used our code with those fixed redshifts to determine the uncertainties on the absolute magnitudes.

The scatter in the TF and FJ mass-luminosity relations are well defined from the observations (see $\S$\ref{sec:tully-fisher} and $\S$\ref{sec:faber-jackson}). However, there is no such scatter for the mass luminosity relation derived from galaxy-galaxy lensing observations, as this relation is the result of a global fit to a halo model where each galaxy provide little information. Therefore, for this relation we assume the same scatter as for the TF relation.

\begin{figure}[h]
\begin{center}
\includegraphics[angle=90.,width=1.0\linewidth]{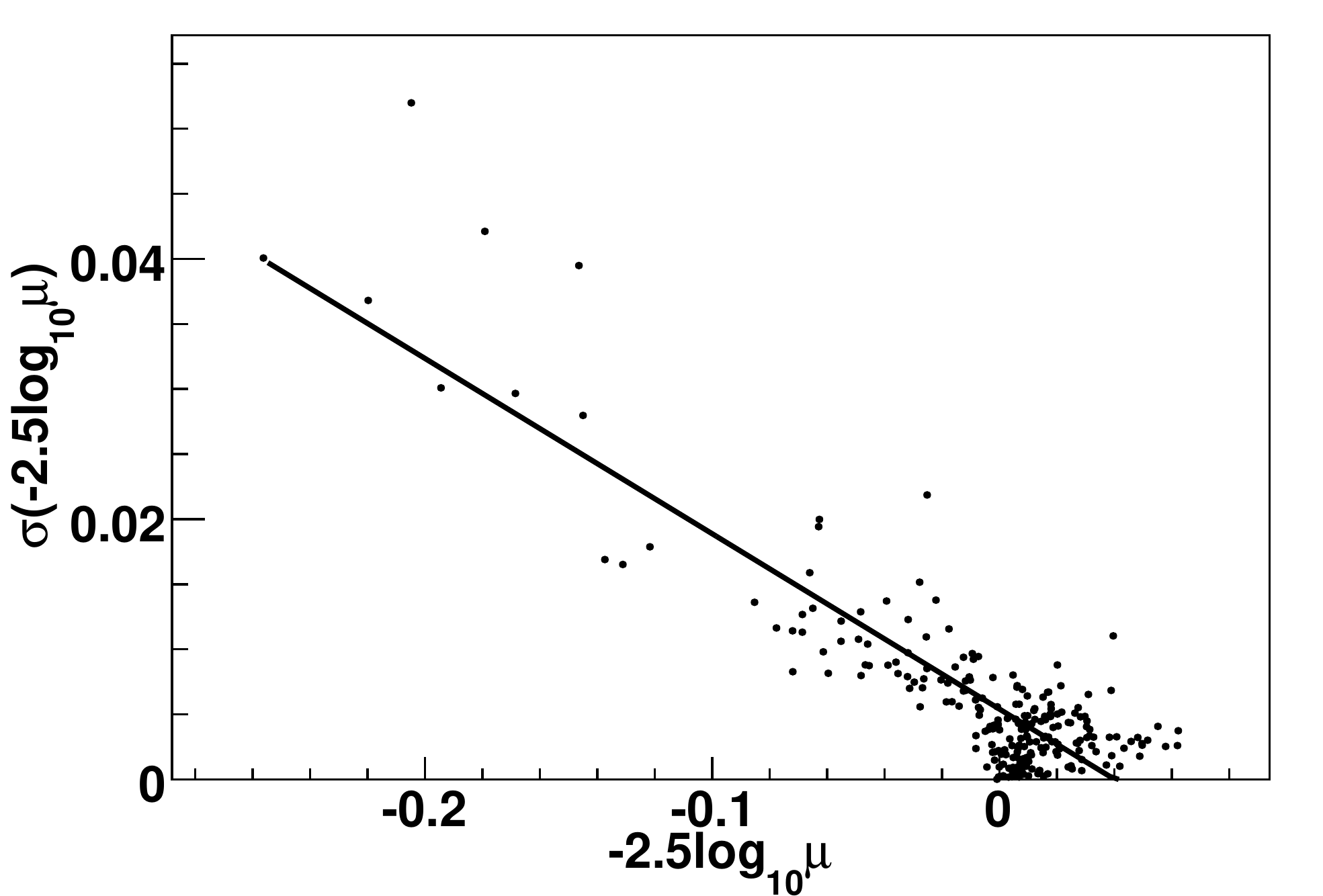}
\caption{Magnification uncertainty as a function of magnification both expressed in magnitudes. The line is a straight line fit to the data. The relative uncertainty is about 17\%.}
\label{erreur_relative}
\end{center}
\end{figure}

In Figure~\ref{erreur_relative} we show the uncertainty of the magnification as a function of the magnification obtained with the TF/FJ relations.  
We obtain a relative uncertainty on the magnification of 17\%. The most
important source of uncertainty comes from the scatter in the
mass-luminosity relation. The uncertainties due to redshift are small
and represent a relative uncertainty of about 5\%.

\section{Signature of the lensing effect}
\label{sec:lensing-signature}
\subsection{Signal discriminant}

As a criterion for a lensing signal detection we have chosen to calculate the weighted correlation coefficient.
\begin{equation}
\rho=\frac{\mathit{cov}(\mu_m,r)}{\sqrt{\mathit{var}(\mu_m)\mathit{var}(r)}}
\end{equation}
with $\mu_m = -2.5 \log_{10}{\mu}$, $\mu$ being the gravitational magnification factor, and $r$ is the residual to the Hubble diagram (hereafter Hubble residual).
The weighted covariance of two variables $x$ and $y$, can be written as :
\begin{equation}
\mathit{cov}(x,y)=\frac{\sum wxy}{\sum w}-\bar{x}\bar{y}
\end{equation}
where $w$ is the weight assigned to each datapoint and $\bar{x}$ and $\bar{y}$ are the weighted means.
Using random lines of sight (ie. random source positions in the galaxy catalog), we have found that weighting with
the inverse of the Hubble residual variance is optimal for
 the signal detection. Note that any scaling applied to $\mu_{m}$ and
  hence any common offset applied to the mass-luminosity relations
   will not change the weighted correlation coefficient and thus not change
    the detection significance.

\subsection{Expected Signal}
\label{sec:simulations-expectations}
Before presenting the results, it is useful to have an idea of what to
expect. We now have all the tools needed to perform detailed Monte
Carlo simulations using the true galaxy catalog and give precise
predictions both for the SNLS 3-year data set and the full SNLS
sample.

Magnification distributions for supernovae at different redshifts were
generated by calculating the magnification factor for a sample of simulated 
supernovae randomly positioned in the true galaxy catalog. In this way it was possible to
simulate the magnification distribution for a sample of supernovae
with a given redshift distribution, chosen as the actual sample
redshift distribution. 
To generate a correlated sample, i.e assuming a correlation
 between the Hubble residual and the magnification, 
we first computed a ``true'' magnification
using the TF/FJ relations applied to galaxies along the line of sight.
We then drew an expected magnification using the uncertainty 
described in \ref{sec:mag_uncertainty_calculation}.
The corresponding Hubble residuals were eventually
generated from the expected magnification and a random offset of r.m.s
0.16 mag in accordance with the SNLS data for which the  Hubble
 residuals exhibit an r.m.s of 0.16 mag. An uncorrelated sample can be 
 generated by randomly assigning a Hubble residual to each expected magnification.

We first simulated a large number of correlated samples similar to the real one with 171
supernovae. We compared the distribution of the weighted correlation
coefficient of these samples with that of uncorrelated samples

\begin{figure}[h]
\begin{center}
\includegraphics[angle=90,width=1.0\linewidth]{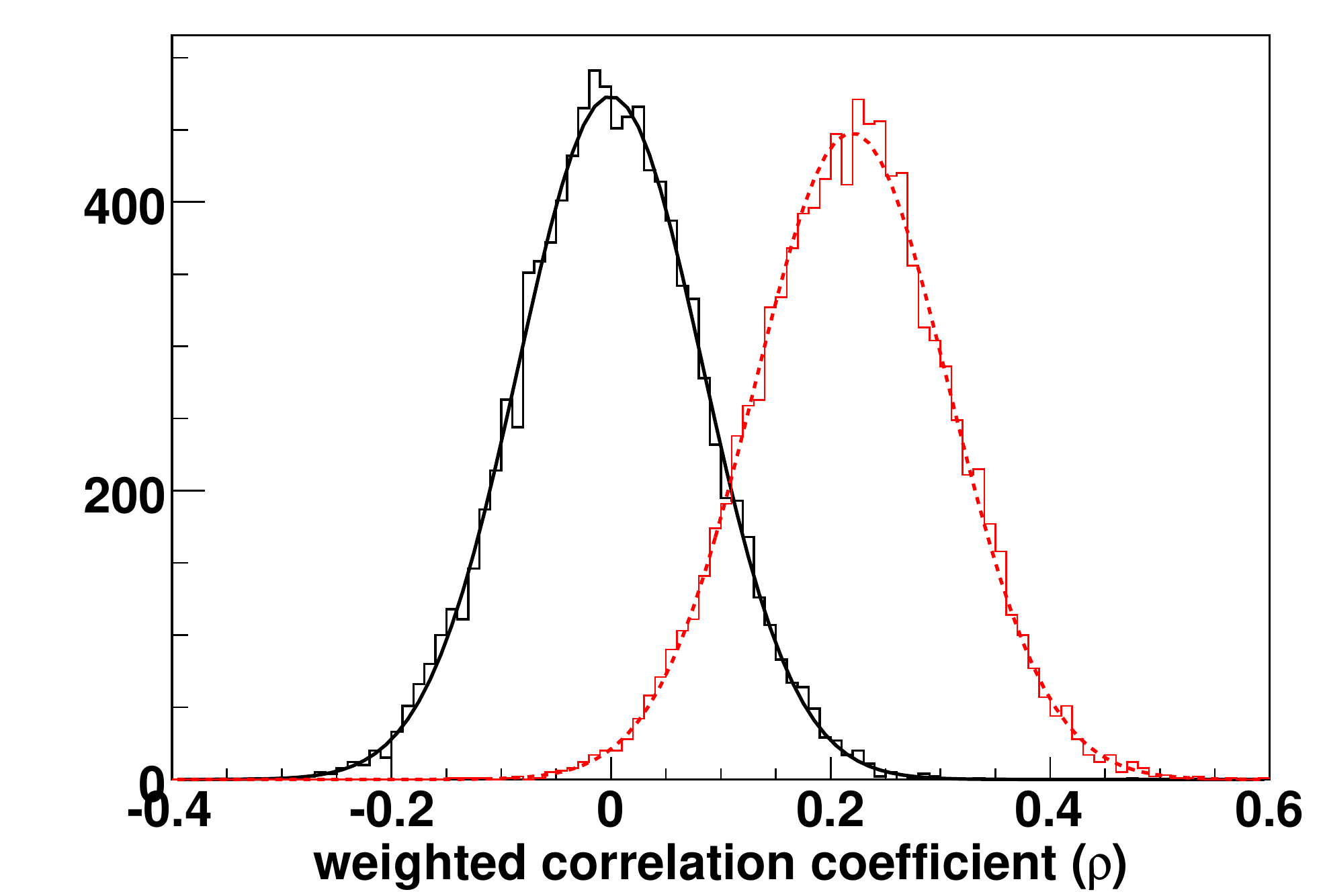}
\caption{Simulated distribution (using real lines of sight) of the weighted correlation coefficients for
correlated samples (dotted) and uncorrelated samples (solid),
for the 3-year SNLS sample.
There is
50 \% probability of finding a 2.5 $\sigma$ significance correlation or better ($\rho=0.22$) and 35 \%
probability of detecting a 3 $\sigma$ signal ($\rho=0.25$).}
\label{mc_rho}
\end{center}
\end{figure}

(see Figure \ref{mc_rho}).  
We find that for the current sample there is 50 \% probability of finding a
2.5~$\sigma$ significance correlation or better and 35 \% probability of detecting a 3~$\sigma$ signal.

\begin{figure}[h]
\begin{center}
\includegraphics[angle=90,width=1.0\linewidth]{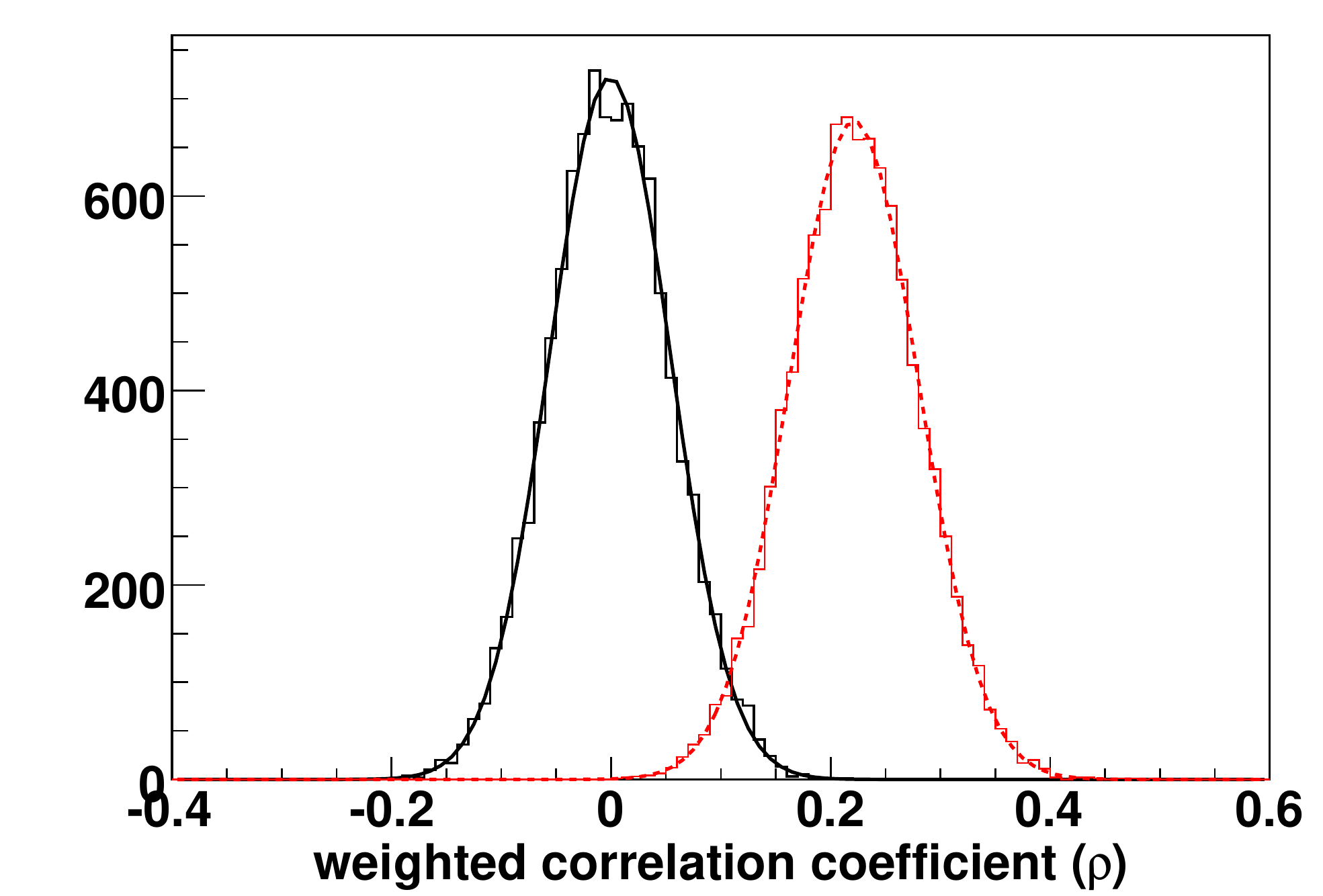}
\caption{Same distributions as Figure \ref{mc_rho}, for the simulated full SNLS sample. There is  80\% probability of detecting a 3 $\sigma$ signal ($\rho=0.22$) or better. }
\label{mc_rho_fullsample}
\end{center}
\end{figure}

For the final SNLS sample we expect $\sim$400
spectroscopically confirmed Type Ia supernovae and $\sim$200 photometrically identified Type Ia
supernovae. Due
to masking, about a third of the supernovae will be rejected. If we do
simulations for 400 supernovae with the same parameters as the current
sample we find that there is 80\% probability of detecting a 3 $\sigma$ signal or more
(see Figure \ref{mc_rho_fullsample} ).

Another question to be addressed is how the errors influence the
possibility of a signal detection. The errors on the magnification are
already small. Monte Carlo simulations performed with various
scaled errors of the magnification support that the scatter in the
magnification has 

little impact on the signal detection. The
signal detection is highly dominated by the scatter in the SN
 Hubble residuals which is not likely to decrease significantly in the near
future. As a consequence improving the signal detection will require better
 statistics and, if possible, higher redshift SNe. 

\subsection{Results}
\label{sec:results}

\subsubsection{The SNLS supernovae}
\label{sec:snls_supernovae}
The 3-year data release contains 233 spectroscopically confirmed
Type Ia supernovae in the redshift range 0.2-1.05 after quality cuts.
 
The Hubble residual of the supernova is the difference in distance modulus of
the supernova and the best cosmology fit. The distance modulus is
estimated from the supernova brightness and other parameters  
resulting from a fit of the SALT2 model \citep{Guy07} 
to the supernova light curves.
The uncertainty of a distance modulus combines three sources: the 
photometric measurement uncertainties, the light curve model scatter
(see \cite{Guy07} for a detailed definition), and the so-called intrinsic
scatter which expresses our lack of complete physical understanding of SN Ia. 
This intrinsic scatter is chosen so that
the minimum $\chi^2$ of cosmological fits to the Hubble diagram matches
the number of degrees of freedom. This ensures that the quoted uncertainties
of Hubble  residuals properly describe their actual scatter.

\subsubsection{Magnification}

\begin{figure}[!h]
\begin{center}
\includegraphics[width=1.\linewidth]{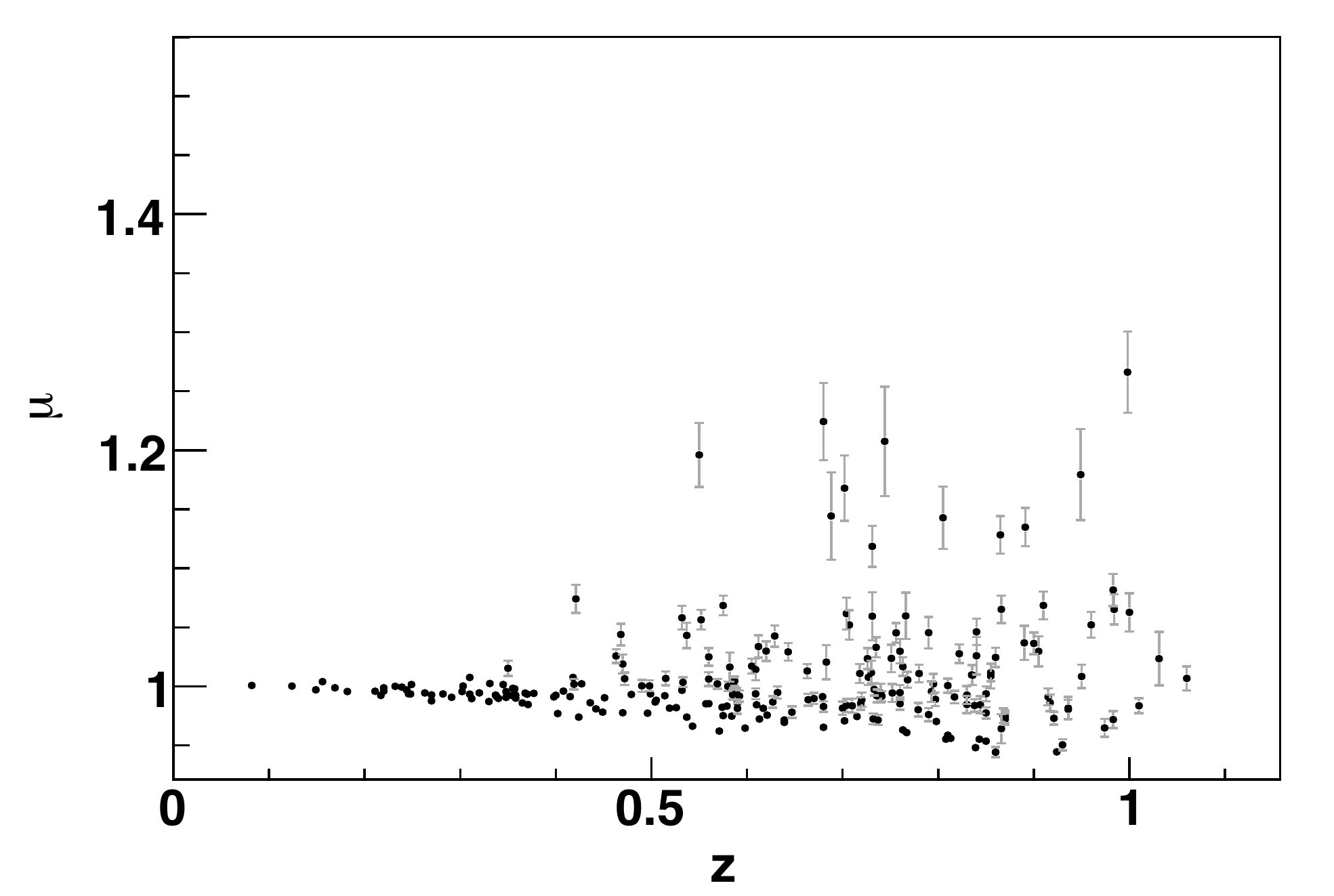}
\includegraphics[width=1.\linewidth]{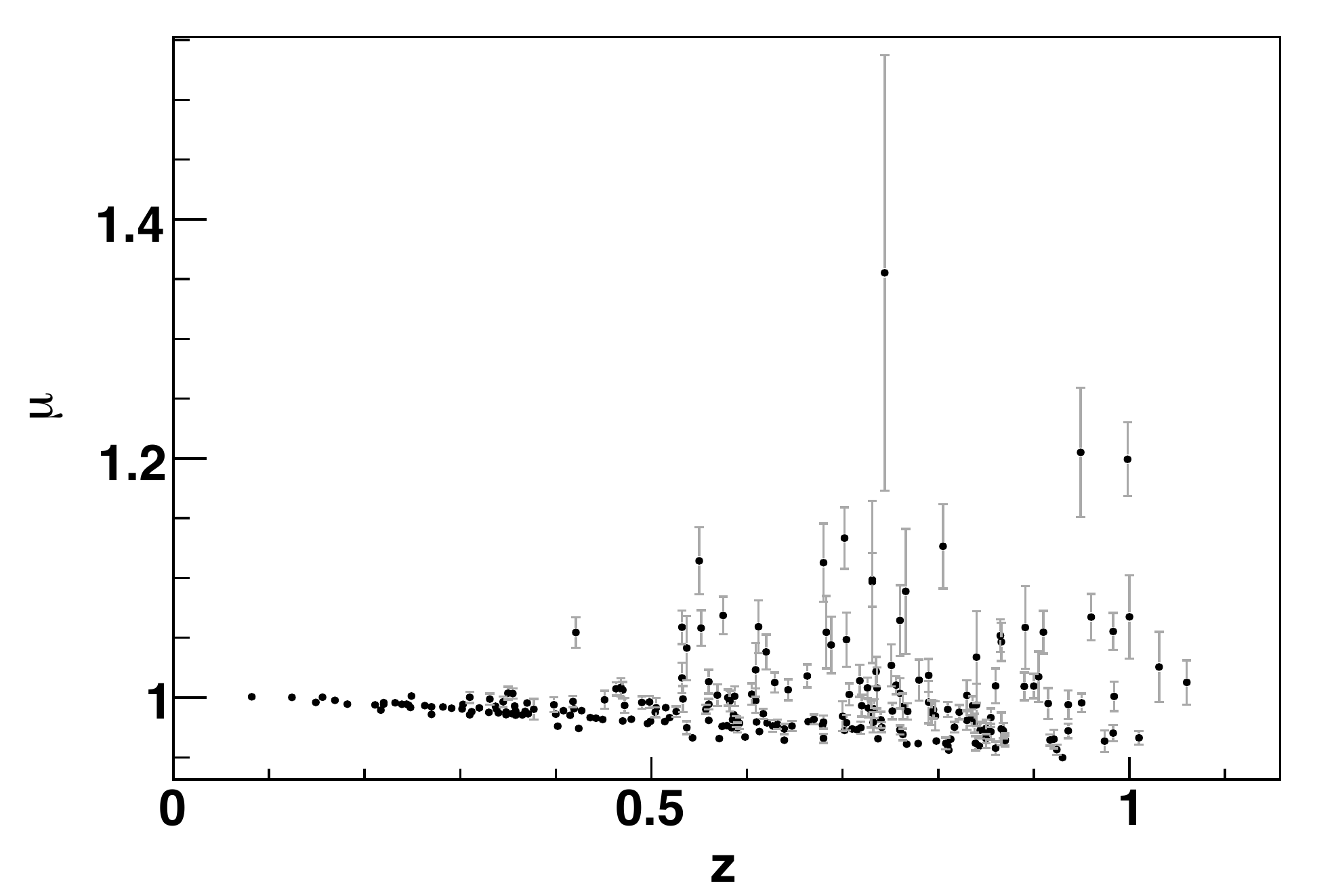}
\caption{Magnification factor of the SNLS supernovae versus redshift. 
Results based on the K06 luminosity-mass relation (top) and on TF and 
FJ relations (bottom). Most of the SNe are slightly demagnified whereas some are significantly magnified.}
\label{mag_z}
\end{center}
\end{figure}

\begin{figure}[!h]
\begin{center}
\includegraphics[angle=90, width=1.\linewidth]{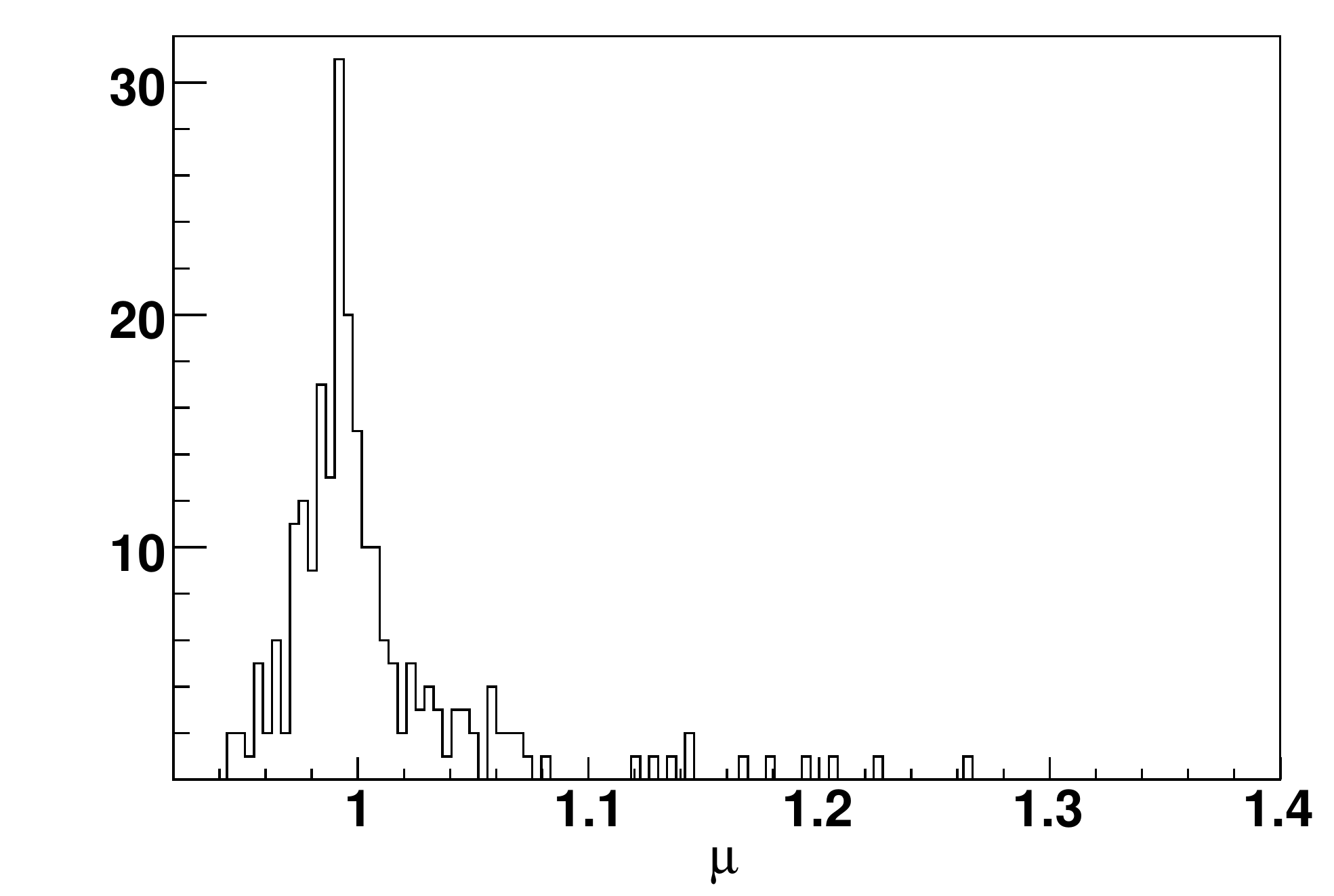}
\includegraphics[angle=90, width=1.\linewidth]{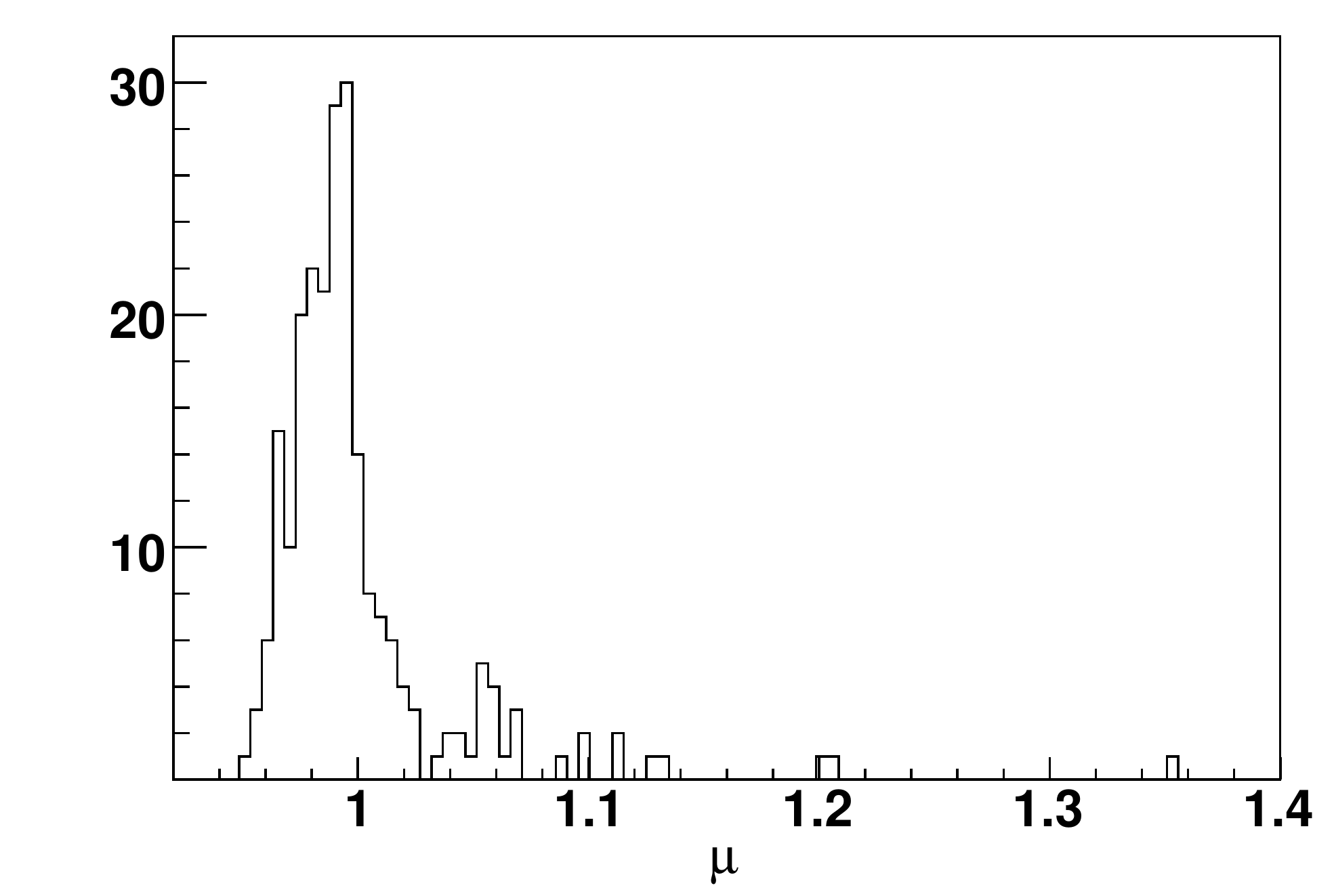}
\caption{Magnification distribution of the SNLS supernovae. Results based on the K06 luminosity-mass relation (top) and on TF and FJ relations (bottom). The magnification distributions peaks at a value slightly below 1 and presents a high magnification tail.}
\label{mag}
\end{center}
\end{figure}

\begin{table*}[!ht]
\begin{center}
\begin{tabular}{||c|c|c|c||c|c|c|c||}
\hline
\multicolumn{2}{||c|}{} & \multicolumn{2}{|c||}{magnification factor $ \mu$} & \multicolumn{4}{|c||}{Most important lensing galaxies}\\
\hline
SN & z & K06 & TF-FJ & z(galaxy) & d (\arcsec) & $\sigma$ km/s (K06) & $\sigma$ km/s  (TF-FJ) \\
\hline
04D1iv  & 0.998 & 1.267$\pm$0.034 & 1.199$\pm$0.031 &0.60 & 7.8& 299 & 295 \\
& & & &  0.51&  5.8 & 217 & 154\\
\hline
04D2kr  & 0.744 & 1.208$\pm$0.046 & 1.355$\pm$0.182& 0.228&1.5 &119 & 150 \\
\hline
05D2by  & 0.891 & 1.135$\pm$0.016 &1.059$\pm$0.035 & 0.66& 1.8& 89 & 50\\
& & & &  0.68&  4.3 & 151 & 167 \\
& & & &  0.44&  2.7 & 88 & 54 \\
\hline
05D2bt  & 0.68 & 1.224$\pm0.33$ & 1.113$\pm$0.033&0.31 &0.5 & 99 & 65 \\
\hline
05D3cx & 0.805 & 1.143 $\pm$0.026 & 1.127$\pm$0.035 & 0.38& 6.4& 224 & 243\\
\hline
03D4cx  & 0.949 & 1.179$\pm$0.038 & 1.205$\pm$0.054&0.45 &5.5 & 246 & 259 \\
\hline
04D4bq  & 0.55  & 1.196$\pm$0.027 & 1.114$\pm$0.028 &0.32 & 4.4&152 & 108  \\ 
& & & &  0.38&  4.1 & 190 & 138 \\
\hline
05D4cq  & 0.702 & 1.168$\pm$0.028 &1.133$\pm$0.026 & 0.28&15.7 & 282 & 298 \\
& & & &  0.43&  2.9 & 106 & 67\\
\hline
\end{tabular}
\end{center}
\caption{The most magnified supernovae and the characteristics of 
the galaxies dominating the magnification.} 
\label{top_20}
\end{table*}

\begin{figure}[!hb]

\begin{center}
\subfigure{
\includegraphics[width=0.95\linewidth]{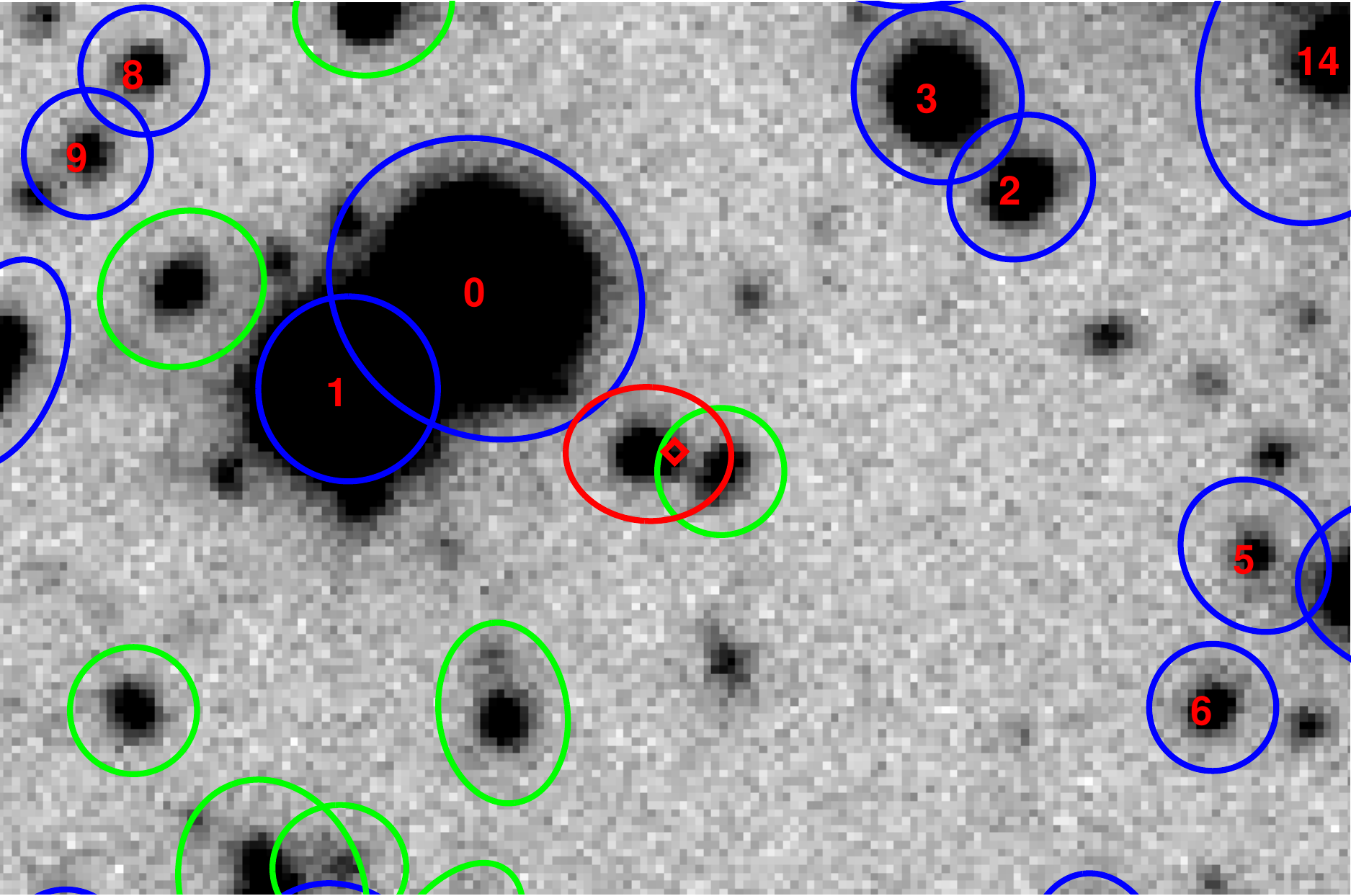}
}
\subfigure{
\includegraphics[width=0.95\linewidth]{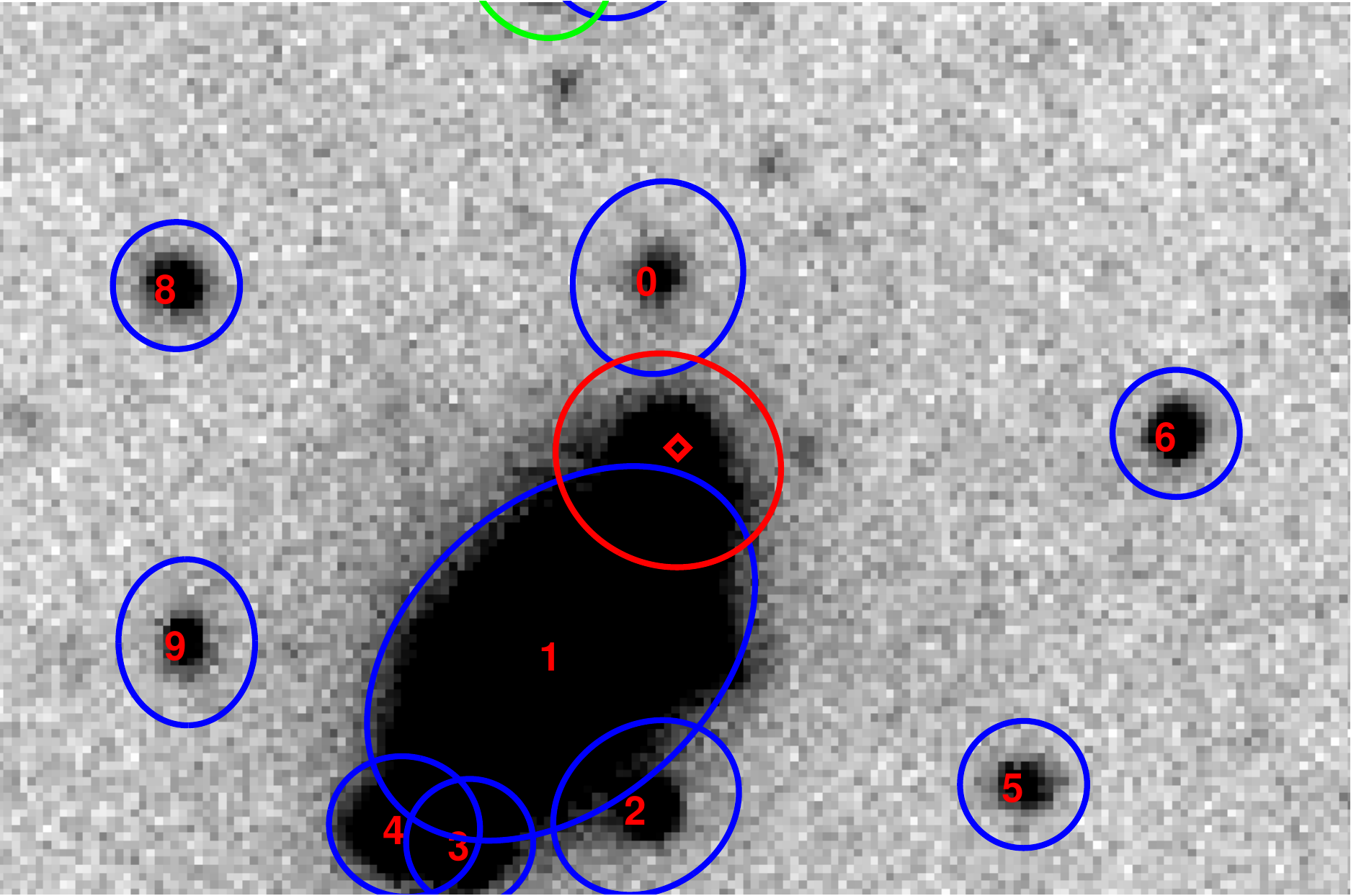}
}
\subfigure{
\includegraphics[width=0.95\linewidth]{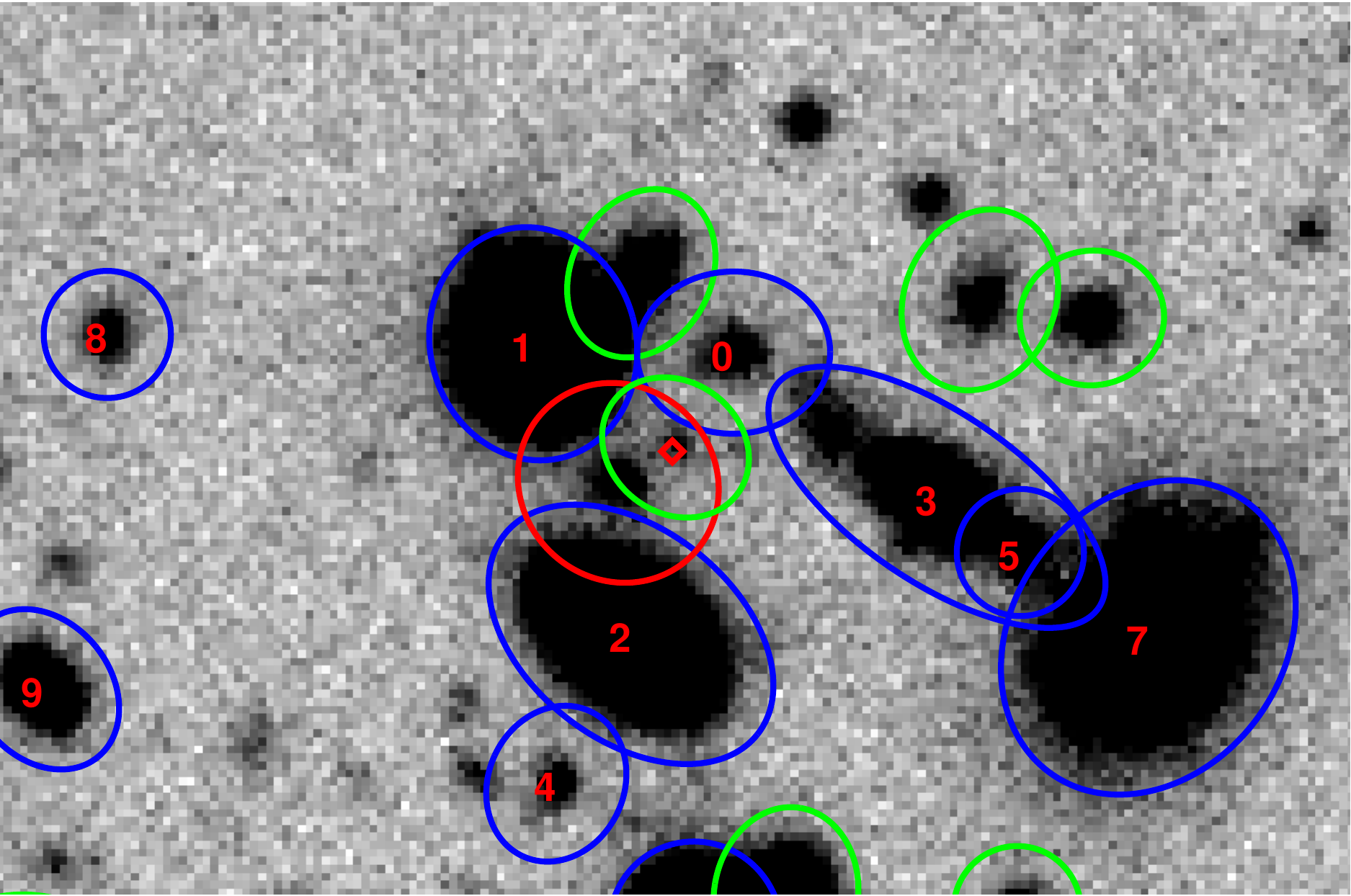}
}
\end{center}

\caption{From the top to the bottom, SN 04D1iv at a redshift of 0.998, SN 03D4cx at a redshift of 0.949 and SN 04D4bq at a redshift of 0.550. In red, the SN and its host. In blue the foreground galaxies labeled according to the closest elliptical distance to the SN and in green the background galaxies. All images span 30\arcsec (horizontally) and are oriented north-up/east-left.}
\label{sn}
\end{figure}

Figure \ref{mag_z} and \ref{mag} show the magnification of each SN as
a function of redshift and the magnification distribution
respectively. As expected, most SNe are demagnified with respect to a
homogeneous universe and some are significantly magnified. Moreover,
the magnification distribution peaks at a value slightly lower than
one and presents a long magnification tail. Images of galaxies along the 
line-of-sight of 3 of the most magnified supernovae (04D1iv, 03D4cx
and 04D4bq) in the 3-year data set are shown in Figure \ref{sn} (see also Figure \ref{04D2kr} \& \ref{05D2bt}). 
All 3 SNe have one or several massive galaxies very close
to the line-of-sight located roughly halfway between the SN and us,
causing the high
magnification. The 8 most magnified supernovae 
along with information on the most important galaxies
causing the magnification are listed in table \ref{top_20}.

\subsubsection{Correlation} 
\label{sec:correlation}
\begin{figure}[!h]
\begin{center}
\includegraphics[angle=90,width=1.\linewidth]{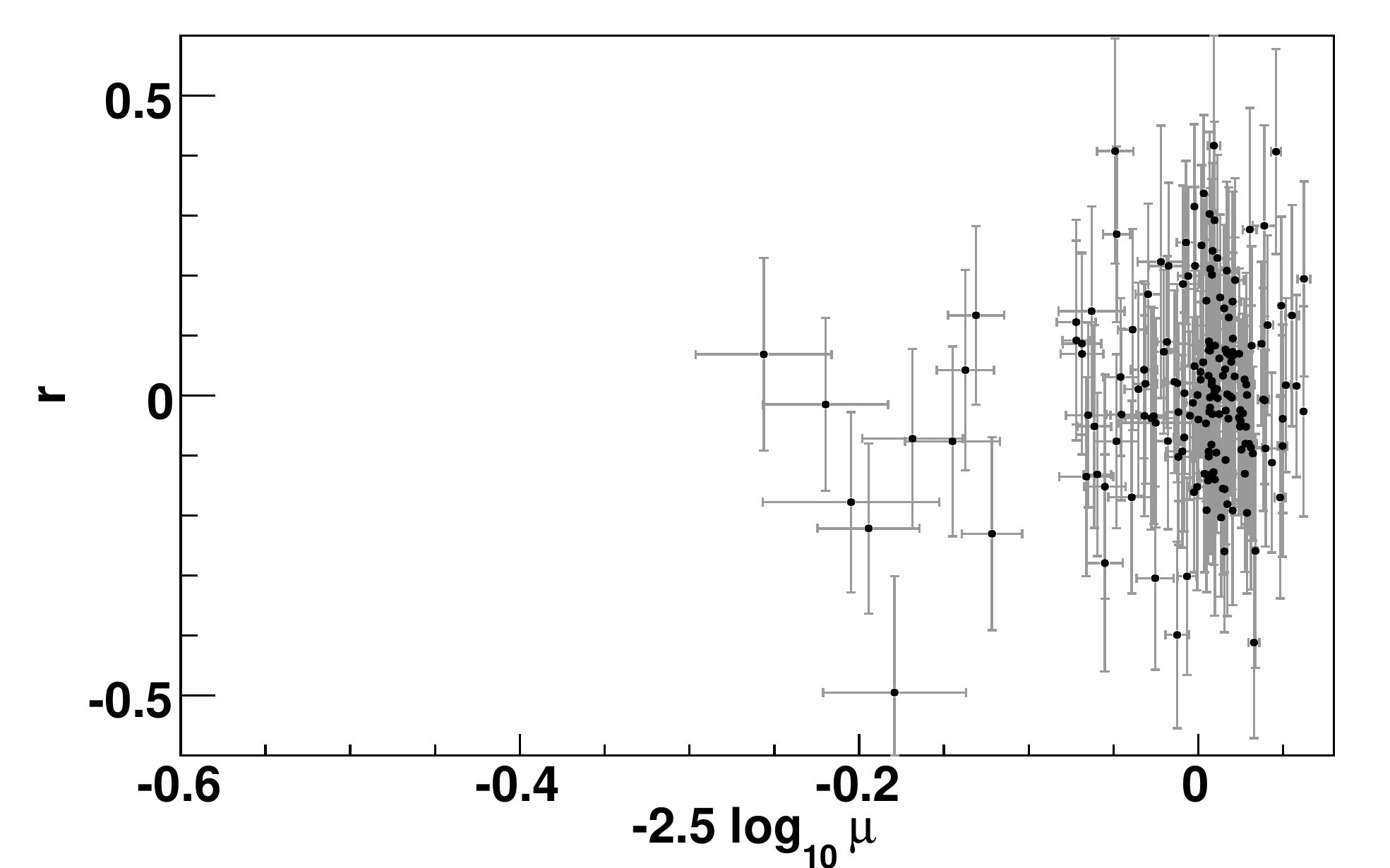}
\includegraphics[angle=90,width=1.\linewidth]{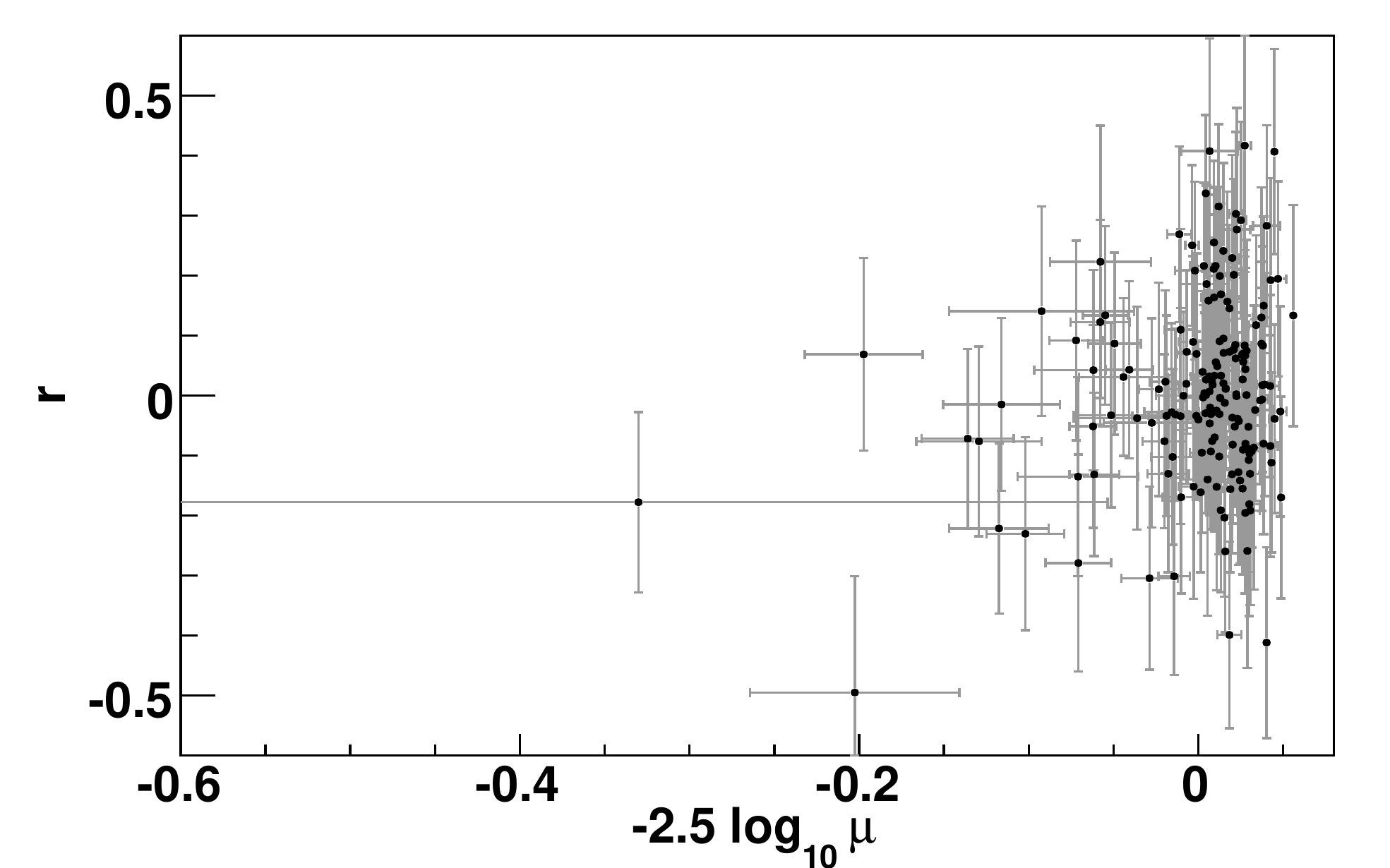}
\caption{The Hubble residuals as a function of magnification for the SNLS. Results based on the K06 luminosity-mass relations (top) and on TF and FJ relations (bottom).}
\label{mag_res_k1}
\end{center}
\end{figure}

We are searching for a correlation between the Hubble residuals from a best fit 
cosmological model and the estimated
magnifications of the supernovae based on foreground galaxy modeling. In
Figure \ref{mag_res_k1} we show a plot of the Hubble residuals of the 171 SNe
versus the estimated magnification.

\begin{figure}[!h]
\begin{center}

\includegraphics[angle=90,width=1.\linewidth]{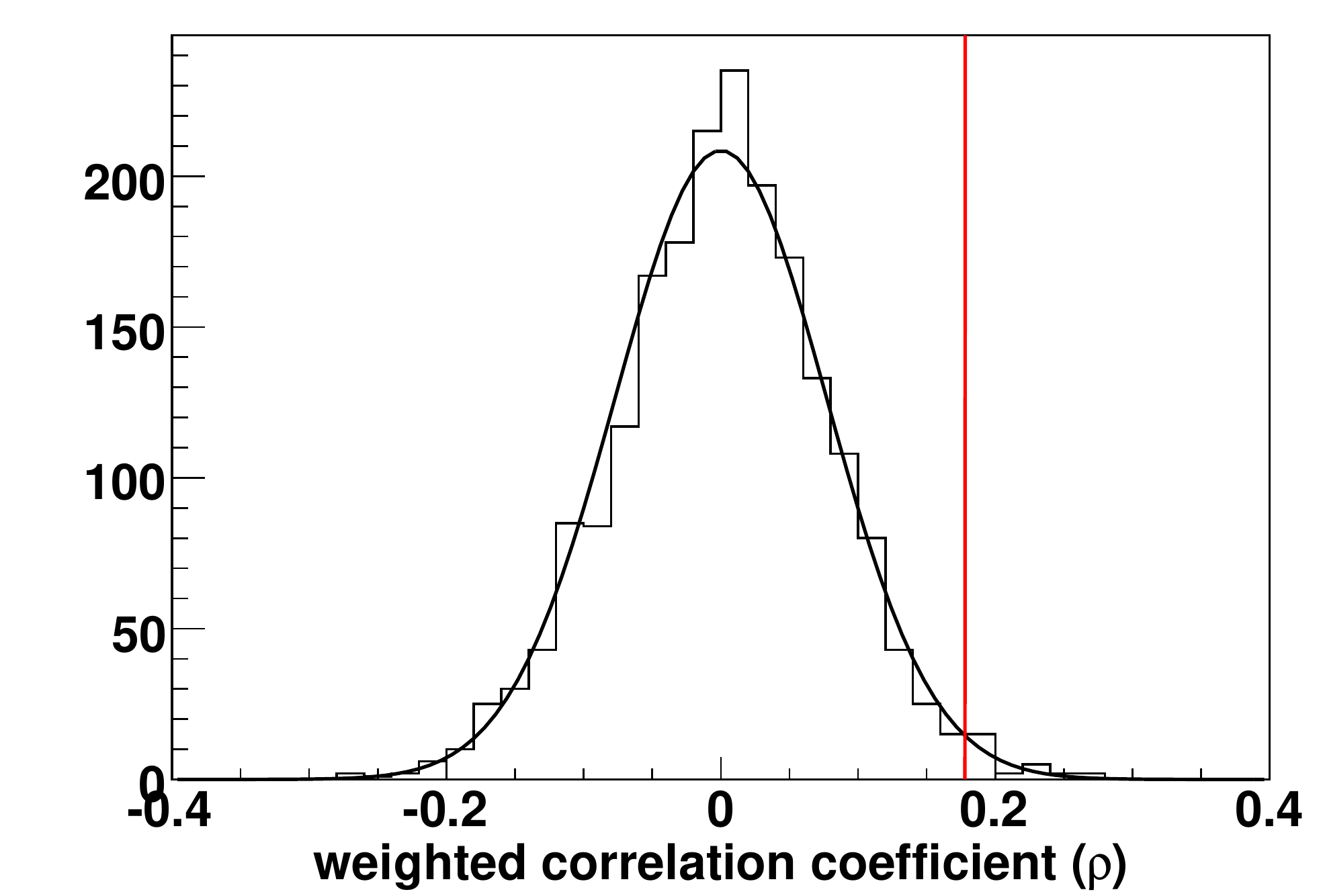}
\caption{
Distribution
of the weighted correlation coefficient for an
uncorrelated sample  when using TF and FJ relations (in black). The red line indicates the value $\rho = 0.18$
obtained for our sample. The measured correlation significance is $2.3 \, \sigma$ (99\% CL).}
\label{rho}
\end{center}
\end{figure}

The weighted correlation coefficient for this sample is $\rho = 0.12$
using the K06 relation and $\rho = 0.18$ using the TF and FJ relations
respectively. To evaluate the strength of the correlation we calculate the
distribution of the weighted correlation coefficient for an
uncorrelated sample and compare it with the obtained value for our
sample (see Figure \ref{rho}). The uncorrelated samples are drawn
by randomly associating Hubble residuals and expected magnifications of the 
real sample. The probability of finding a larger weighted correlation
coefficient than the measured one from an uncorrelated sample is 5\% 
using the K06 relation and 1\% using the TF and FJ relations, corresponding
to 1.6 and $2.3\,\sigma$ detections respectively.

It is tempting to attribute the stronger detection of the TF/FJ
relation to the fact that this method considers separately elliptical
and spiral galaxies. In order to test the influence of the separation,
we run the analysis again, but with a random galaxy type assignment
(however preserving the measured proportions). We then find that the
significance of the detection drops from 2.3 to 1.4 $\sigma$, and
conclude that the higher significance of our detection using TF/FJ
relations can be attributed to the distinction between spiral and
elliptical galaxies, rather than chance. So, our primary result is the
detection of supernovae lensing using TF/FJ relations at the 99\% CL.
Note that simulations assuming a perfect galaxy typing (see section \ref{sec:simulations-expectations}) 
give rise to a mean detection level of 2.5 $\sigma$ (see Figure \ref{mc_rho}). 
The 2.3 $\sigma$ detection we find is thus close to the optimal 
so that further improvements on galaxy typing should not give rise 
to a much larger significance of the signal.

In order to test the scale of galaxy mass estimates, we fit the
slope $a$ relating Hubble residuals and expected magnifications
 : $<r> = a < \mu_m>$ and find $a=0.65~\pm~0.30$. Hence the data
 are consistent with the TF/FJ mass-luminosity relations 
at the 1.2 $\sigma$ level, with a precision of 30\%.

\begin{figure}[!h]
\begin{center}
\includegraphics[angle=90,width=0.95\linewidth]{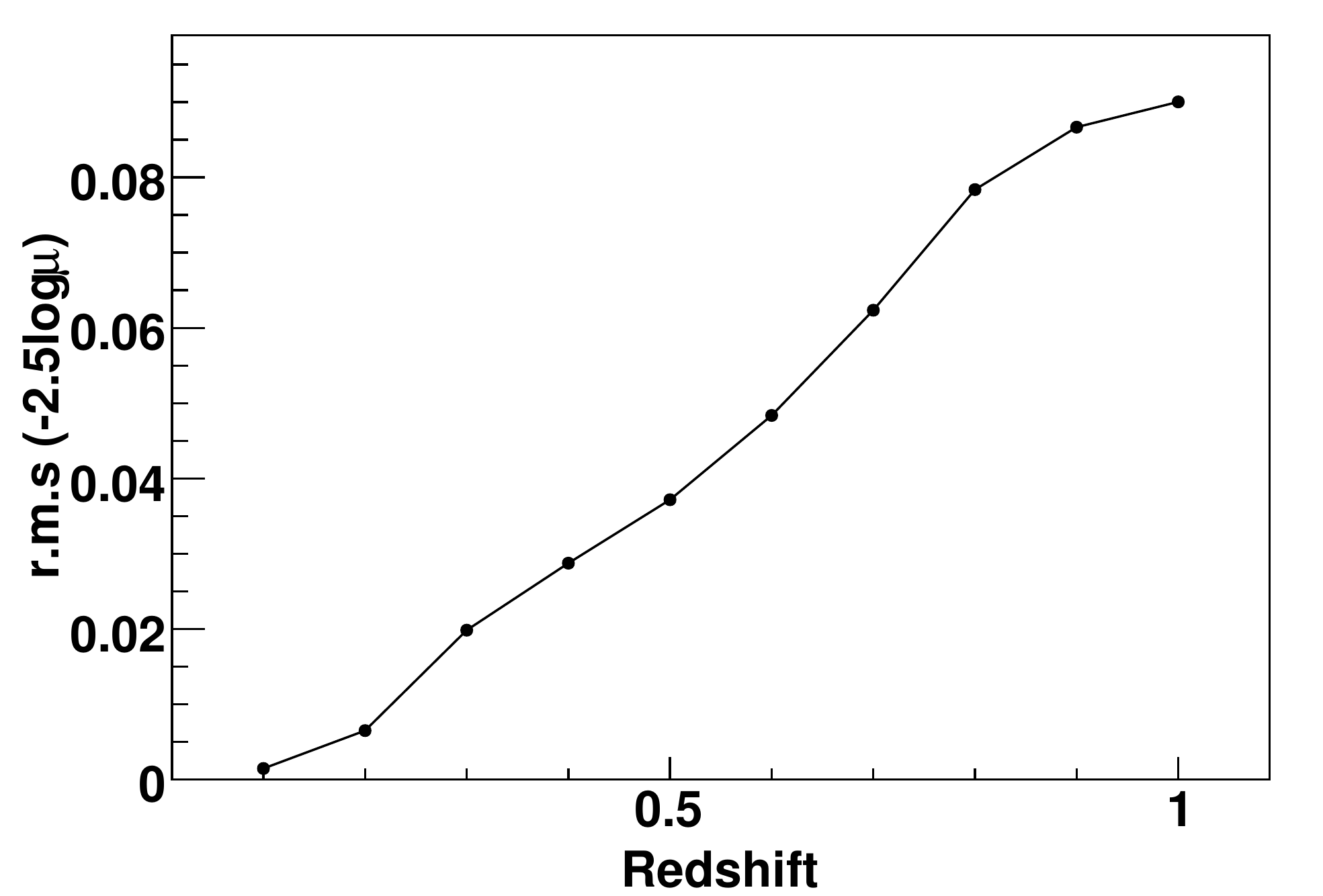}
\caption{
Expected magnification scatter from random lines of sight
as a function of the source redshift, using the TF/FJ 
mass-luminosity relations.}
\label{fig:magnification_rms}
\end{center}
\end{figure}

Using random lines of sight in the real data, we can estimate the
increase of Hubble diagram scatter expected from gravitational
lensing, as a function of redshift.  This is shown in Figure
\ref{fig:magnification_rms} for the TF/FJ relations, which can be
roughly described as $\sigma(\mu_m) = 0.08 \times z$. Alternatively, if
 we use the value of $a$ derived from a fit of the relation between Hubble residuals and magnifications, we obtain a lower value $\sigma(\mu_m) = (0.05 \pm 0.022) \times z$.

\section{Conclusion}
\label{sec:conclusion}

We have calculated the expected magnification of the SNLS 3-year
sample from the foreground galaxy properties and searched for a
correlation with residuals from the Hubble diagram. A correlation is
detected at the 99\% CL, compatible with a slope of 1.

The expected magnifications cause an extra scatter in the Hubble diagram
approximated by $0.08 \times z$ from the TF/FJ relations we used, which becomes $(0.05 \pm 0.022) \times z $ 
once these TF/FJ relations are calibrated with the supernova data. We show that separating 
the galaxy sample into a blue and a red population based on a $U\!-\!V$ 
color cut increases the significance of the detection. This is due to the fact that the mass estimate 
and hence the induced magnification is significantly different for a red and a blue galaxy of same luminosity (the TF and FJ relations).
Simulations also
point to the fact that a signal detection is dominated by the
number of SNe, their redshift distribution and the scatter in the SN
residuals. Reducing the scatter in the estimated magnification by
increasing the photometric redshift precision or reducing the scatter in the TF and FJ relations
 have little effect on the probability of a signal detection.

Finally, simulations using the true galaxy catalog show that using the full
SNLS data set ($\sim$400 expected spectroscopically confirmed Type Ia SNe
and $\sim$200 photometrically identified Type Ia SNe) there is 80\%
chance of detecting a $3 \, \sigma$ signal or more.  \\

\begin{acknowledgements}
This article is based on the work of a Ph.D. thesis. TK acknowledges
the support of the French ministry of Education and Research 
and the Dark Cosmology Centre funded by the Danish
National Research Foundation. We thank O. Ilbert for providing us
with high resolution photometric redshifts in the COSMOS field prior
to publication. We also thank Michel Fioc, Damien Le Borgne and Jens Hjort for useful discussions. This
research has made use of the NASA/ IPAC Infrared Science Archive,
which is operated by the Jet Propulsion Laboratory, California
Institute of Technology, under contract with the National Aeronautics
and Space Administration. The data reduction was carried out
at the IN2P3 computing center in Lyon, France. 
\end{acknowledgements}
\bibliographystyle{aa_like_apj}

\bibliography{biblio}
\end{document}